\documentclass[twocolumn]{aastex63}
\usepackage{bm}
\usepackage{url}
\usepackage{amsmath}
\usepackage{color}
\usepackage{graphicx}
\usepackage{hyperref}
\usepackage[T1]{fontenc}

\received{}
\revised{}
\accepted{}
\submitjournal{ApJ}

\shorttitle{Dust characterization of the HL Tau disk}
\shortauthors{Ueda et al.}
\begin{document}

\title{
Multi-Wavelength Dust Characterization of the HL Tau Disk and Implications for Planet Formation
}

\correspondingauthor{Takahiro Ueda}
\email{takahiro.ueda@cfa.harvard.edu}

\author[0000-0003-4902-222X]{Takahiro Ueda}
\affil{Center for Astrophysics, Harvard \& Smithsonian, 60 Garden Street, Cambridge, MA 02138, USA}

\author[0000-0003-2253-2270]{Sean M. Andrews}
\affil{Center for Astrophysics, Harvard \& Smithsonian, 60 Garden Street, Cambridge, MA 02138, USA}

\author[0000-0003-2862-5363]{Carlos Carrasco-Gonz\'alez}
\affiliation{Instituto de Radioastronom\'{\i}a y Astrof\'{\i}sica (IRyA), Universidad Nacional Aut\'onoma de M\'exico (UNAM)}

\author[0000-0003-4753-8759]{Osmar M. Guerra-Alvarado}
\affiliation{Leiden Observatory, Leiden University, PO Box 9513, 2300 RA Leiden, The Netherlands}

\author[0000-0002-1886-0880]{Satoshi Okuzumi}
\affiliation{Department of Earth and Planetary Sciences, Institute of Science Tokyo, Meguro, Tokyo 152-8551, Japan}

\author[0000-0003-1451-6836]{Ryo Tazaki}
\affiliation{Department of Earth Science and Astronomy, Graduate School of Arts and Sciences, The University of Tokyo, 3-8-1 Komaba, Meguro, Tokyo 153-8902, Japan}

\author[0000-0003-4562-4119]{Akimasa Kataoka}
\affiliation{National Astronomical Observatory of Japan, Osawa 2-21-1, Mitaka, Tokyo 181-8588, Japan}



\begin{abstract}
We present a comprehensive analysis of the HL Tau dust disk by modeling its intensity profiles across six wavelengths (0.45 to 7.9 mm) with a resolution of 0$\farcs$05 ($\sim7$ au).
Using a Markov Chain Monte Carlo (MCMC) approach, we constrain key dust properties including temperature, surface density, maximum grain size, composition, filling factor, and size distribution.
The full fitting, with all parameters free, shows a preference for organics-rich dust with a low filling factor in the outer region ($r \gtrsim 40$ au), where the spectral index is $\sim3.7$, but amorphous-carbon-rich dust also reasonably reproduces the observed intensity profiles.
Considering the scattering polarization observed at 0.87 mm, compact, amorphous-carbon-rich dust is unlikely, and moderately porous dust is favored.
Beyond 40 au, the maximum dust size is likely $\sim100~{\rm \mu m}$ if dust is compact or amorphous-carbon rich. 
However, if the dust is moderately porous and organics-rich, both the predicted dust surface density and dust size can be sufficiently large for the pebble accretion rate to reach $\sim10M_{\oplus}~{\rm Myr^{-1}}$ in most regions, suggesting that pebble accretion could be a key mechanism for forming planets in the disk.
In contrast, if the dust is amorphous-carbon-rich, forming a giant planet core via pebble accretion is unlikely due to the combined effects of low dust surface density and small dust size required to match the observed emission, suggesting other mechanisms, such as disk fragmentation due to gravitational instability, may be responsible for planet formation in the HL Tau disk.

\end{abstract}


\section{Introduction}
Planet formation begins with the collisional evolution of dust particles in protoplanetary disks orbiting protostars. Dust particles grow into larger aggregates, fragment into smaller ones, or drift radially toward the central star, depending on their size \citep{BM93,Weidenschilling77,BW08,Birnstiel+10,Wada+13,Birnstiel24}. 
The dust radial drift plays an important role in planetesimal formation via local dust accumulation followed by the streaming instability \citep{YG05}, as well as in the subsequent evolution of protoplanets through pebble accretion \citep{OK10,LJ12}.
Observational constraints on dust mass and size thus are essential for understanding planet formation.

Dust properties are inferred from the spectral behavior of thermal emission at (sub-)millimeter to centimeter wavelengths (see \citealt{Andrews20} for a review). 
If optical depths are sufficiently lower than unity at a given observing frequency $\nu$, the disk's total flux $F_{\nu}$ scales with dust mass $M_{d}$, dust opacity $\kappa_{\nu}$ and the Planck function $B_{\nu}(T)$; $F_{\nu}\propto M_{d}\kappa_{\nu}B_{\nu}(T)$ (e.g., \citealt{Hildebrand83,Beckwith+90,AW+05}).
Therefore, the spectral index $\alpha$ of the total flux, $F_{\nu}\propto \nu^{\alpha}$, is related to the opacity index $\beta$ ($\kappa_{\nu}\propto \nu^{\beta}$) as $\alpha=\beta+2$ in the Rayleigh-Jeans limit, providing information about the dust properties. 
Conversely, in the optically thick regime, the emission follows the Planck function with modification by scattering \citep{MN93,Liu19,Zhu+19,SL20,Ueda+20}.

Recent developments with the Atacama Large Millimeter/submillimeter Array (ALMA) have enabled multi-wavelength dust characterization in spatially resolved disk observations \citep{Carrasco-Gonzalez+19,Macias+21,Sierra+21,Ueda+22,Guidi+22,Houge+24,Guerra-Alvarado+24}.
Most previous studies have focused on deriving fundamental dust properties, such as the maximum dust radius $a_{\rm max}$, surface density $\Sigma_{\rm d}$, and temperature $T$, by adopting a specific dust model to relate $a_{\rm max}$ to optical properties like extinction opacity and albedo. 
However, the underlying dust model in disks remains highly uncertain (e.g., \citealt{Birnstiel+18}).

Dust composition is a key factor in determining opacities and thus plays a crucial role in interpreting millimeter observations. 
The dust composition has been inferred based on observations of protoplanetary disks and the interstellar medium, as well as from the solar system comets (e.g., \citealt{MRN77,Pollack+94,D'Alessio+01,Zubko+04,Patzold+16}).
Carbonaceous materials are thought to constitute a large fraction of the refractory core of dust particles, but it remains unclear in what form they are present.
The model proposed by the Disk Substructures at High Angular Resolution Project (DSHARP) includes refractory organics as the carbonaceous component, accounting for 40\% of the total dust mass \citep{Birnstiel+18}. 
By contrast, the dust model of \citet{Ricci+10} employs amorphous carbon as the carbonaceous material, which leads to significantly higher absorption opacity than in the DSHARP model.
The disk population synthesis studies tend to favor such absorbent dust, because dust growth and radial drift make the disk optically thin quickly, which is inconsistent with observations \citep{Stadler+22,Delussu+24}.
In reality, the abundances of refractory organics and amorphous carbon may vary spatially within the disk due to thermal processing and chemical reactions (e.g., \citealt{Lee+10,CS12,GT17}).

Dust porosity plays a crucial role in dust evolution. Micron-sized dust particles coagulate to form fluffy aggregates with a filling factor of $\ll0.1$ \citep{BW00,KB04,Suyama+08,Krijt+15,Lorek+18}.
The high porosity promotes the collisional growth of dust \citep{Ormel+07,Okuzumi+12,Kataoka+13,GG20}.
Recent near-infrared observations suggest the presence of highly porous dust in the disk surface \citep{Ginski+23,Tazaki+23}.
On the other hand, the dust porosity may decrease via compression due to gas pressure and filling of voids by smaller dust grains produced by fragmentation, resulting in a moderate porosity with on the order of $\sim0.1$ \citep{Kataoka+13a, Dominik+16,Tanaka+23,Michoulier+24}.
The moderately porous dust is commonly detected by observations of small bodies in the Solar System (e.g., \citealt{Guttler+19}).
Although previous multi-wavelength millimeter modeling of Stokes $I$ emission has attempted to constrain the porosity, no clear consensus has been established (\citealt{Guidi+22,Guerra-Alvarado+24}).
Meanwhile, recent millimeter polarization measurements indicate that moderately porous dust, with a filling factor of $\sim0.1$, is preferred to explain scattering-induced polarization over a broad wavelength range \citep{Zhang+23, Lin+24, Ueda+24}.

The protoplanetary disk around HL Tau is among the brightest and most extensively observed disks, making it an ideal target for detailed investigations of dust properties.
Multi-wavelength analysis using four observing wavelengths ranging from 0.9 to 8 mm suggested dust sizes between 0.5 and 1.5 mm throughout the disk \citep{Carrasco-Gonzalez+19}.
However, adding 0.45 mm data indicate larger dust in the inner regions and smaller grains in the outer regions compared to those inferred by \citet{Carrasco-Gonzalez+19} \citep{Guerra-Alvarado+24}.
Multi-wavelength polarimetric observations of HL Tau support the moderately porous dust scenario \citep{Zhang+23, Lin+24}.
While these previous studies have made several assumptions about the dust model, the rich observational data of the HL Tau disk potentially enable us to constrain the dust properties without relying on such assumptions.

In this paper, we investigate the dust properties in the HL Tau disk using the MCMC method to better constrain key parameters and assess their uncertainties. 
Our analysis considers not only the maximum dust size, surface density, and temperature but also composition, porosity, and size distribution. 
In Section \ref{sec:obs}, we describe the observational data used in our analysis. 
Section \ref{sec:method} outlines our method, and Section \ref{sec:result} presents our results. 
We discuss our findings in Section \ref{sec:discussion} and conclude in Section \ref{sec:summary}.

\section{Observations} \label{sec:obs}

\begin{figure*}[tbh]
\begin{center}
\includegraphics[width=2\columnwidth]{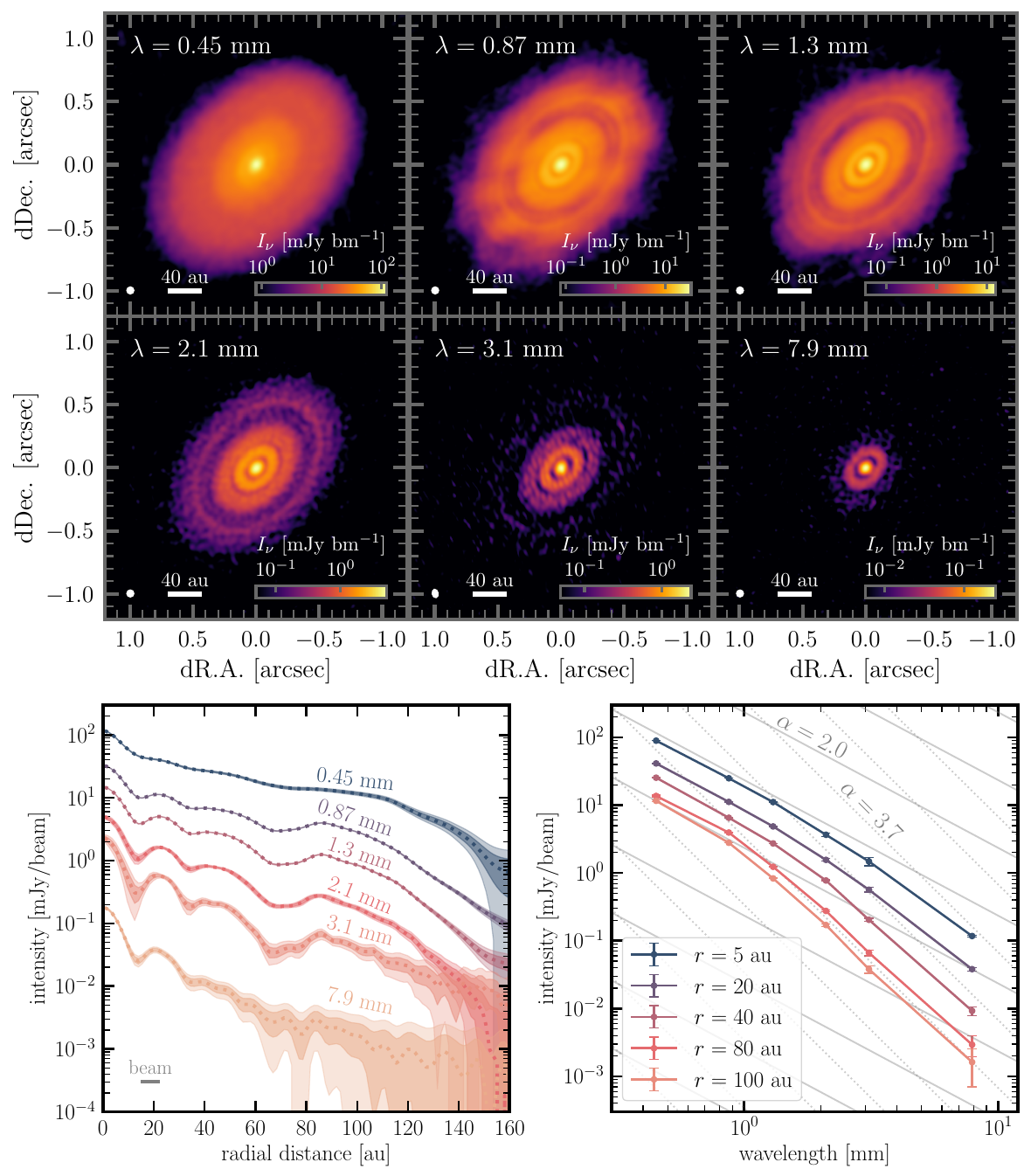}
\caption{
Top: images of dust continuum emission of the HL Tau disk. An arcsinh stretch is applied to the color scale.
Bottom left: azimuthally-averaged radial intensity profiles.
The thin and thick transparent regions denote the 1$\sigma$ and 2$\sigma$ RMS uncertainty, respectively.
The beam size is 0$\farcs$061$\times$0$\farcs$041 at 3.1 mm and 0$\farcs$05$\times$0$\farcs$05 at other wavelengths (with equal beam areas).
Bottom right: Spectral behavior of the intensity at $r = 5$, 20, 40, 80, and 100 au.
The errorbars represent the 1$\sigma$ RMS uncertainties.
The gray solid and dotted lines denote spectral slopes of 2 and 3.7, respectively.
}
\label{fig:HLTau}
\end{center}
\end{figure*}

We model the observed radial intensity profile of the HL Tau disk taken at six wavelengths, $\lambda=0.45$ (ALMA Band 9), 0.87 (Band 7), 1.3 (Band 6), 2.1 (Band 4), 3.1 (Band 3) and 7.9 mm (VLA Ka$+$Q), as shown in Figure \ref{fig:HLTau}.
The 0.45 mm data are originally from \cite{Guerra-Alvarado+24} and the data at 0.87, 1.3 and 7.9 mm are from \cite{Carrasco-Gonzalez+19}. 
These datasets have a common angular resolution of $0\farcs05$, corresponding to a spatial resolution of 7.35 au for the distance of HL Tau (147 pc; \citealt{Galli+18}).

For the 2.1 mm data,  the ALMA Science Verification data have been used in the literature \citep{Carrasco-Gonzalez+19,Guerra-Alvarado+24}.
However, those original data lack short baselines ($\lesssim170~{\rm m}$), so spatial filtering results in an artificially depressed intensity distribution, particularly in the outer regions ($\gtrsim60$ au; see Appendix \ref{sec:B4}).
To remedy this, we use the short-baseline data from the ALMA archive (ID: 2019.1.00134.S, PI: Ian Stephens) that includes baselines down to $\sim30$ m.
The short-baseline data are reduced and calibrated using the Common Astronomical Software Application ({\tt CASA}) package \citep{CASA}.
The initial flagging of the visibilities and calibrations for the bandpass characteristics, complex gain, and flux scaling are performed using the pipeline scripts provided by ALMA.
After that, two rounds of phase-only calibration and one round of phase and amplitude calibration are applied with solution intervals of 60 s, 30 s and infinity (scan lengths), respectively.

The calibrated short-baseline data is concatenated with the long-baseline data which is self-calibrated by using the pipeline scripts provided by ALMA.
As the two data sets show slight misalignment in the phase center, we apply {\tt phaseshift} and {\tt fixplanets} to the datasets before the concatenation to ensure that the both datasets have the same phase center (following \citealt{Andrews+18}).
The concatenated data are again self-calibrated with two rounds of phase-only and one round of phase and amplitude calibration.
The calibrated data are imaged by {\tt tclean} with adopting Briggs weighting with a robust parameter of $-1.5$.
The signal-to-noise ratio in our data is sufficiently high to support such a weighting without significantly degrading image fidelity.
The obtained image has the FWHM beam size of 0$\farcs$044$\times$0$\farcs$036.
The {\tt imsmooth} function is applied to match the beam size to the common resolution of 0$\farcs$05$\times$0$\farcs$05, allowing for accurate intensity comparisons on the same spatial scale.

For the 3.1 mm data, we use the data taken and calibrated in the ALMA Science Verification program \citep{ALMAPartnership15}.
The calibrated data are imaged by {\tt tclean} with adopting Briggs weighting with a robust parameter of $-1.5$.
The resulting angular resolution is 0$\farcs$06$\times$0$\farcs$035, which is slightly better than the other data but is elongated.
The image is smoothed using {\tt imsmooth} to achieve a beam size of 0$\farcs$061$\times$0$\farcs$041, resulting in a beam area equivalent to 0$\farcs$05$\times$0$\farcs$05.

Figure \ref{fig:HLTau} shows the spectral variation of the observed intensity at specific radial locations (see Appendix \ref{sec:spectralindex} for detailed radial profiles). 
The spectral index follows $\alpha \sim 2$ at $r \lesssim 40$ au for wavelengths $\lambda \lesssim 1.3$ mm, while it is steeper at $\lambda \gtrsim 1.3$ mm. 
In the outer regions ($r = 80$ and $100$ au), the spectral index remains $\sim 2$ at $\lambda \lesssim 0.87$ mm but increases to $\sim 3.7$ at $\lambda \gtrsim 0.87$ mm.
Overall, the primary difference from the previous dataset is the inclusion of Band 3 data and the correction of Band 4 data for missing flux due to the lack of short-baseline coverage.

\section{Methods} \label{sec:method}

\subsection{Analytical model of the intensities}
The observed radial intensity profiles are fitted with an analytical model to deduce dust properties.
The axisymmetry of the HL Tau disk allows for azimuthal averaging of the intensity, which improves the signal-to-noise ratio and enables one-dimensional radial analysis.
The analytical intensity at a given frequency is a function of three variables -- dust temperature $T$, extinction optical depth $\tau_{\nu}$, and effective albedo $\omega^{\rm eff}_{\nu}$ (albedo corrected for the forward scattering with asymmetry parameter; \citealt{HG41}) -- and is described as (e.g., \citealt{Sierra+24})
\begin{eqnarray}
I_{\rm \nu}=B_{\rm \nu}(T)\left\{ 1-\exp{\left(-\frac{\tau_{\nu}}{\mu}\right)} +\omega^{\rm eff}_{\nu} F(\tau_{\nu},\omega^{\rm eff}_{\nu})\right\},
\label{eq:intensity}
\end{eqnarray}
where $B_{\nu}(T)$ is the Planck function, and $\mu\equiv\cos{i}$ represents the effect of the disk inclination.
The function $F$ represents the effect of scattering:
\begin{eqnarray}
F(\tau_{\nu},\omega^{\rm eff}_{\nu})=\frac{f_{1}(\tau_{\nu},\omega^{\rm eff}_{\nu})+f_{2}(\tau_{\nu},\omega^{\rm eff}_{\nu})}{\exp{(-\sqrt{3}\epsilon_{\nu}\tau_{\nu})}(\epsilon_{\nu}-1)-(\epsilon_{\nu}+1)}
\label{eq:Ffactor}
\end{eqnarray}
where
\begin{eqnarray}
f_{1}(\tau_{\nu},\omega^{\rm eff}_{\nu})=\frac{ 1-\exp{\left\{ -(\sqrt{3}\epsilon_{\nu}+1/\mu)\tau_{\nu}\right\}} }{\sqrt{3}\epsilon_{\nu}\mu+1}
\end{eqnarray}
and
\begin{eqnarray}
f_{2}(\tau_{\nu},\omega^{\rm eff}_{\nu})=\frac{\exp{\left(-\tau_{\nu}/\mu\right)}-\exp{(-\sqrt{3}\epsilon_{\nu}\tau_{\nu})}}{\sqrt{3}\epsilon_{\nu}\mu-1},
\end{eqnarray}
with $\epsilon_{\nu}\equiv\sqrt{1-\omega^{\rm eff}_{\nu}}$.
Since the extinction optical depth is the product of the extinction opacity $\kappa_{\rm ext, \nu}$ and the dust surface density $\Sigma_{\rm d}$, Equation \eqref{eq:intensity} effectively includes four variables ($T$, $\kappa_{\rm ext,\nu}$, $\Sigma_{\rm d}$ and $\omega_{\rm \nu}^{\rm eff}$).
For a given dust model (e.g., composition, filling factor $f_{\rm fill}$, and power-law index of the size distribution $p_{\rm d}$), the extinction opacity and effective albedo are characterized by $a_{\rm max}$.
Therefore, the most commonly deduced variables from multi-wavelength analysis are $T$, $\Sigma_{\rm d}$ and $a_{\rm max}$ (e.g., \citealt{Carrasco-Gonzalez+19,Guidi+22,Ueda+22}).

\subsection{Opacities}

\begin{figure*}[tbh]
\begin{center}
\includegraphics[width=2\columnwidth]{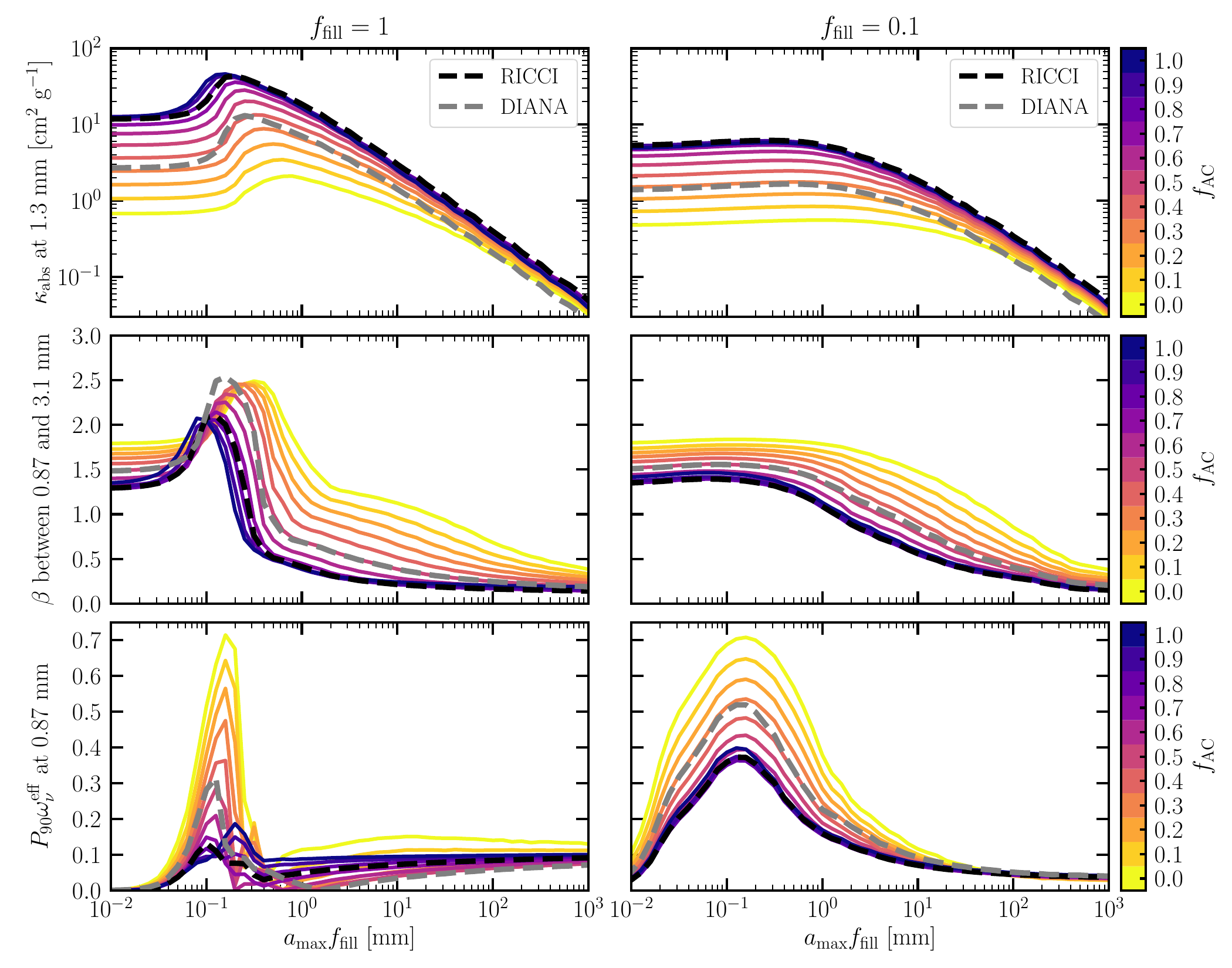}
\caption{
Optical properties of our dust model with $f_{\rm fill}=1$ (left) and 0.1 (right).
Top: absorption opacity at 1.3 mm. 
Middle: opacity index between 0.87 and 3.1 mm.
Bottom: polarization efficiency at 0.87 mm.
The power-law index of dust-size distribution is set to 3.0.
The black and gray dashed lines denote opacities calculated with dust composition used in \citet{Ricci+10} and \cite{Woitke+16}, respectively.
Although the original DIANA model assumes $f_{\rm fill}=0.75$, we set $f_{\rm fill}=1$ or 0.1 for fair comparison.
}
\label{fig:opac}
\end{center}
\end{figure*}

The dust composition, porosity, and size distribution in disks have large uncertainties, and no consensus has been reached regarding them (e.g., \citealt{Birnstiel+18}).
Therefore, in this work, we treat them as fitting parameters in addition to $T$, $\Sigma_{\rm d}$ and $a_{\rm max}$, resulting in a total of six fitting parameters.

To parameterize the dust composition, we introduce a parameter $f_{\rm AC}$ that represents the fraction of amorphous carbon in the carbonaceous material in the dust.
Our dust composition model is based on the DSHARP model \citep{Birnstiel+18} which is a mixture of water ice (20\% in mass; \citealt{WB08}), astronomical silicates (33\%; \citealt{Draine03}), troilite (7\%; \citealt{HS96}) and carbonaceous material (40\%).
The original DSHARP model assumes refractory organics \citep{HS96} as the carbonaceous material.
However, the exact form of the carbonaceous material is uncertain and has a significant impact on the opacities.
Particularly, presence of amorphous carbon significantly enhances dust absorbency \citep{Zubko+96}.
Therefore, we introduce $f_{\rm AC}$ to describe the fraction of amorphous carbon within the carbonaceous material.
For instance, $f_{\rm AC}=0.2$ means that 20\% of the refractory organics of the DSHARP dust are replaced by amorphous carbon.

Figure \ref{fig:opac} shows the optical properties of our dust model calculated with {\tt OpTool} \citep{Dominik+21}.
The irregularity of dust shape is considered by using the distribution of hollow spheres method with $f_{\rm max}=0.8$, where $f_{\rm max}$ describes the maximum fraction of the hollow \citep{Min+05}.
The porosity is implemented using vacuum as an additional material, which is mixed with the other materials using the Bruggeman rule.
The horizontal axis in Figure \ref{fig:opac} is $a_{\rm max}f_{\rm fill}$ instead of $a_{\rm max}$, because this product represents the column density (mass-to-area ratio) of an individual grain and better characterizes its optical (and also aerodynamical) properties than $a_{\rm max}$ \citep{Kataoka+14}.
One can see that the absorption opacity is one order of magnitude higher for $f_{\rm AC}=1$ than for $f_{\rm AC}=0$ (DSHARP model).
Including porosity (lowering the filling factor) suppresses the opacity enhancement at $a_{\rm max}f_{\rm fill}\sim\lambda/2\pi$ due to Mie interference \citep{BH98}.
The strong peak in opacity slope $\beta$ associated with this opacity enhancement also disappears in porous dust.

Figure \ref{fig:opac} (bottom) shows the polarization efficiency, the product of polarization degree at $90^{\circ}$ scattering ($P_{90}$) and effective albedo.
The polarization efficiency is maximized at $a_{\rm max} f_{\rm fill} \sim \lambda/2\pi$ \citep{Kataoka+15}, but the width of the peak at $ a_{\rm max} f_{\rm fill} \sim \lambda/2\pi$ is much broader for moderately porous dust ($f_{\rm fill} \sim 0.1$) compared to compact dust.
More porous dust ($f_{\rm fill}\ll0.1$) shows a small polarization efficiency \citep{Tazaki+19}.
For a given filling factor, more amorphous-carbon-rich dust exhibits lower polarization efficiency due to its higher absorbency (i.e., lower albedo). 

In millimeter studies, opacity models proposed by \citet{Ricci+10} (hereafter RICCI) and \citet{Woitke+16} (DIANA) have been also widely used. 
The RICCI model assumes a mixture of water, silicate, and amorphous carbon in a 3:1:2 volume ratio (0.2857:0.3417:0.3726 by mass), while the DIANA model assumes pyroxene and amorphous carbon in a 4:1 ratio (0.87:0.13 by mass). 
The RICCI and DIANA models exhibit absorption opacities roughly 10 and 3 times higher, respectively, than our model with $f_{\rm AC}=0$ (DSHARP model) and resemble our models with $f_{\rm AC}\sim0.9$--1 and $f_{\rm AC}\sim0.3$--0.4. 
This mainly reflects the amorphous-carbon mass fractions in the RICCI (0.37) and DIANA (0.13) models, which correspond to replacing 92.5\% or 32.5\% of the carbonaceous material in the DSHARP model with amorphous carbon. 
Although other materials also influence the optical properties, adjusting the amorphous-carbon fraction in the DSHARP model largely reproduces the behavior of both models.

\subsection{Markov Chain Monte Carlo fitting}
We analyze the observed intensity at each radius to infer the six parameters: dust temperature, surface density, maximum dust size, amorphous carbon fraction, filling factor and power-law of size distribution.
We compute the posterior probability distribution of our model at each radius using the MCMC implementation within the {\tt emcee} package \citep{Foreman-Mackey+13}.
The posterior probability distributions are calculated using a standard likelihood function:
\begin{eqnarray}
P(I_{\rm obs}|T,\Sigma_{\rm d},a_{\rm max},f_{\rm AC},f_{\rm fill},p_{\rm d}) \propto \exp{\left(-\frac{\chi^{2}}{2}\right)},
\end{eqnarray}
with
\begin{eqnarray}
\chi^{2}=\sum_{\nu}\left(\frac{I_{{\rm obs},\nu}-I_{{\rm model},\nu}}{\sigma_{\nu}}\right)^{2},
\end{eqnarray}
where $I_{{\rm obs},\nu}$ and $I_{{\rm model},\nu}$ are the observed and model intensity at frequency $\nu$, respectively.
The uncertainty in the observed intensity at frequency $\nu$, $\sigma_{\nu}$, is given as
\begin{eqnarray}
\sigma_{\nu}^{2} = \Delta I_{{\rm obs},\nu}^{2} + (\delta_{\rm \nu} I_{{\rm obs},\nu})^{2},
\end{eqnarray}
where $\Delta I_{{\rm obs},\nu}$ is the standard deviation of the azimuthally averaged intensity at frequency $\nu$ and $\delta_{\nu}$ represents the absolute flux calibration uncertainty ($1\sigma$) at frequency $\nu$.
We set $\delta_{\nu}$ as 10\% at 0.45 mm, 5\% at 0.87, 1.3 and 7.9 mm, and 2.5\% at 2.1 and 3.1 mm, following the ALMA and VLA official values.

For the dust temperature, we use a prior based on the temperature inferred from the luminosity of HL Tau (e.g., \citealt{Dullemond+01}); 
\begin{eqnarray}
T_{\rm model}(r,L_{*},\phi)=\left( \frac{\phi L_{*}}{8\pi r^{2}\sigma_{\rm SB}} \right)^{1/4},
\label{eq:temp}
\end{eqnarray}
where $\phi$ is a grazing angle, $\sigma_{\rm SB}$ is the Stefan-Boltzmann constant, and $L_{*}$ is the stellar luminosity.
The luminosity is assumed to be a Gaussian distribution centered at $L_{*}=6L_{\odot}$ with a standard deviation of $\sigma_{L}=5L_{\odot}$ that cover the typical estimates from literature \citep{Men'shchikov+99,Pinte+16,Liu+17}.
The grazing angle $\phi$ is set to be uniformly distributed from 0.01 to 0.3.
Based on these, the prior for the dust temperature is given as 
\begin{equation}
p_{1}(r,T)\propto \int \exp\left\{-\frac{1}{2} \left( \frac{L_{\rm model}(r,T,\phi)-L_{\rm *}}{\sigma_{L}} \right)^{2} \right\} {\rm d}\phi,
\label{eq:t_prior}
\end{equation}
where $L_{\rm model}$ is the inverse function of $T_{\rm model}$ with respect to $L_{*}$.
We apply uniform priors for the other parameters in our fiducial model, but also adopt priors for $a_{\rm max}, f_{\rm AC}, f_{\rm fill}$ and $p_{\rm d}$ based on the polarization efficiency at 0.87 mm (see Section \ref{sec:pol}).

The fitting parameters are sampled within the following ranges: 
$T$ between 5 and 200 K, 
$\log{\Sigma_{\rm d}}$ [${\rm g~cm^{-2}}$] between $-4$ and $1.5$,
$\log{a_{\rm max}}$ [${\rm mm}$] between $-2$ and $3$, 
$f_{\rm AC}$ between 0 and 1,
$\log{f_{\rm fill}}$ between $-2$ and 0, and
$p_{\rm d}$ between 2.5 and 4.
The MCMC simulation is performed using 180 walkers and 9000 steps, with the initial 30\% of the steps discarded as the burn-in period to ensure convergence. 
An initial sampling is conducted by running the MCMC for 2700 steps, with the initial conditions for each walker randomly assigned within the parameter ranges. 
The initial positions of the walkers for the main MCMC run are randomly chosen from the 67\% confidence interval of the posterior distributions obtained from this initial sampling.

\section{Results} \label{sec:result}
In this section, we present the MCMC analysis results for the six-wavelength observations of the HL Tau disk.  
Section \ref{sec:full} shows the analysis with all parameters free,  
Section~\ref{sec:fix} presents cases with fixed dust composition and filling factor,  
and Section~\ref{sec:pol} includes results incorporating polarization-based priors.

\subsection{Full fitting} \label{sec:full}

\begin{figure*}[tbh]
\begin{center}
\includegraphics[width=2\columnwidth]{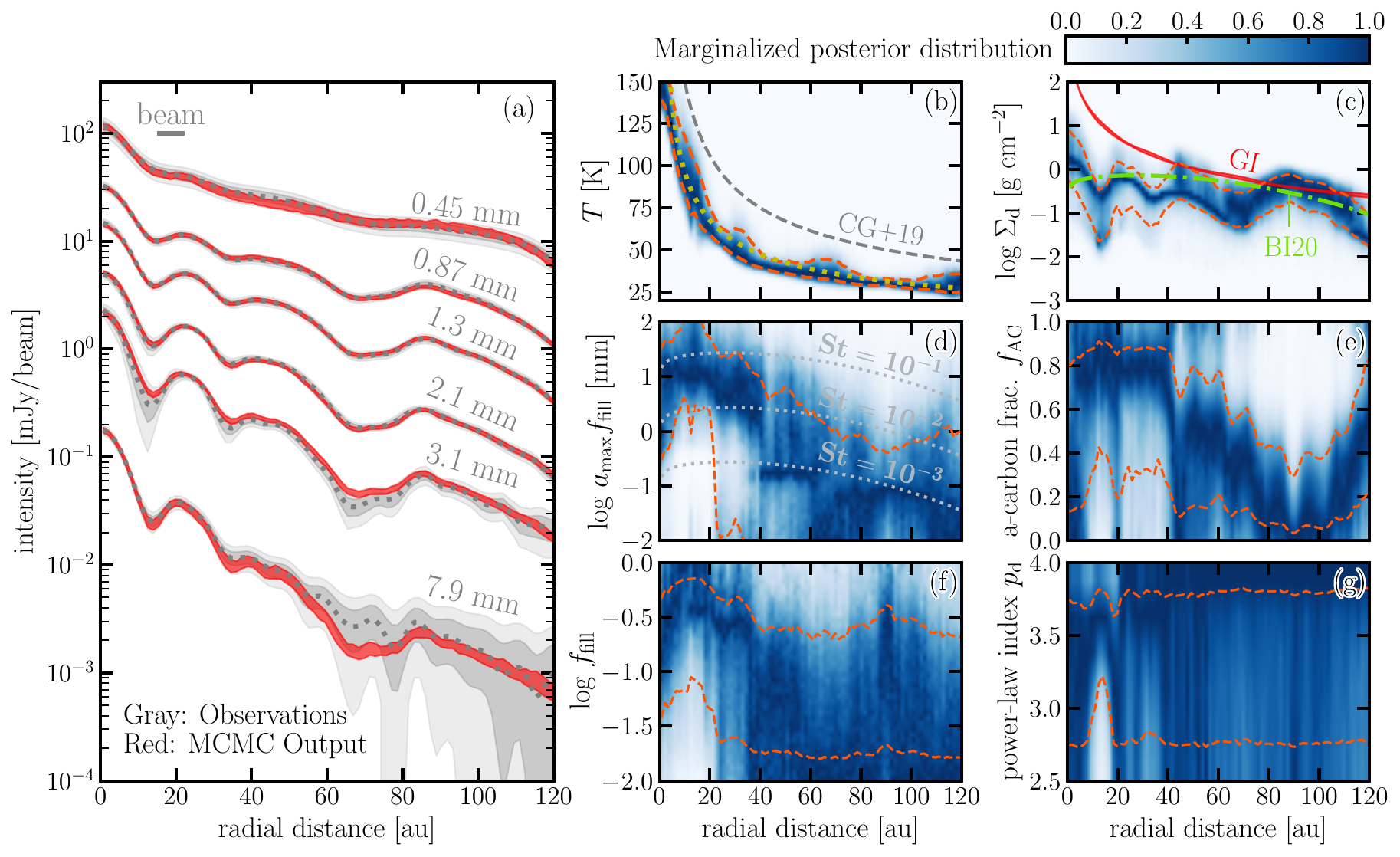}
\caption{
MCMC fitting results with all parameters free.
Left: comparison of observed intensities (gray) with MCMC-derived models (red), computed using the 68\% confidence interval of the posterior distributions.
Right two columns: marginal posterior distributions of all parameters. 
The orange dashed lines in all posterior plots denote the 68\% confidence interval.
In panel~(b), the yellow dotted line indicates the power-law temperature profile $T = 55\,(r/30~\mathrm{au})^{-0.51}~\mathrm{K}$ (Equation~\ref{eq:temp_simple}), while the gray dashed line shows the temperature profile $T = 86.6\,(r/30~\mathrm{au})^{-0.5}~\mathrm{K}$ from \citet{Carrasco-Gonzalez+19}.
In panel~(c), the red-shaded region marks the gravitational instability threshold (Equation~\ref{eq:GI}), estimated from the 68\% confidence range of the dust temperature posterior. 
The green dash-dotted line represents 1\% of the gas surface density inferred from ${\rm ^{13}C^{17}O}$ observations (Equation~\ref{eq:sigmag}).
In panel~(d), the gray dotted lines indicate Stokes numbers ${\rm St} = 10^{-1}$, $10^{-2}$, and $10^{-3}$. 
}
\label{fig:full}
\end{center}
\end{figure*}

Figure~\ref{fig:full} presents the marginalized posterior distributions of the six fitting parameters from the MCMC analysis and the predicted intensity profiles within the 68\% confidence interval.
Our model successfully reproduces the observed intensity profile within $\sim1\sigma$ level.
The dust temperature posterior is well constrained owing to the Band 9 observations \citep{Guerra-Alvarado+24}, at which the disk is optically thick and the albedo is low, and a parametric fit in the form $T_{30} (r/{\rm 30~au})^{-q}$ for the 68\% confidence interval yields
\begin{eqnarray}
T=55(r/{\rm 30~au})^{-0.51}~{\rm K},
\label{eq:temp_simple}
\end{eqnarray}
consistent with the analytical temperature model (Equation \ref{eq:temp}) with $\phi=0.12$.
The derived dust temperature is $\sim36$\% lower than that derived by \citet{Carrasco-Gonzalez+19}, $T = 86.6\,(r/30~\mathrm{au})^{-0.5}~\mathrm{K}$.
This difference is due to the fact that, in \citet{Carrasco-Gonzalez+19}, millimeter-sized dust grains with high albedo are favored, which requires a higher dust temperature to explain the observed brightness.

Compared to the dust temperature, the other parameters exhibit relatively large dispersions in their posterior distributions.
The dust surface density exhibits a flat profile (with substructures) ranging from $\sim0.1$ to $\sim1$ ${\rm g\,cm^{-2}}$, with an uncertainty of one order of magnitude.
As discussed later (see Section \ref{sec:fix}), this flat surface density profile is due to the predicted amorphous carbon fraction that increases toward the inner regions; higher amorphous carbon fraction (higher absorption opacity) in the inner regions results in lower dust surface density.
The predicted dust surface density is close to that of a gravitationally unstable disk;
\begin{eqnarray}
\Sigma_{\rm d,GI} = f_{\rm d2g}\frac{c_{\rm s}\Omega_{\rm K}}{\pi G Q},
\label{eq:GI}
\end{eqnarray}
where $f_{\rm d2g}$ is the dust-to-gas mass ratio, $Q$ is Toomre's parameter \citep{Toomre64}, and $\Omega_{\rm K}=\sqrt{GM_{*}/r^{3}}$ is the Keplerian frequency.
We adopt $f_{\rm d2g}=0.01$, $Q=1.4$ and $M_{*}=2.1M_{\odot}$ \citep{Yen+19}.
The sound speed $c_{\rm s}$ is calculated using the 68\% confidence interval of the dust temperature posterior. 
Between 50 and 100 au, the $Q=1.4$ curve aligns with 0.01 times the gas surface density  derived from the ${\rm ^{13}C^{17}O}$ observations \citep{BI20};
\begin{eqnarray}
\Sigma_{\rm g} = \Sigma_{0}\left( \frac{r}{r_{\rm c}} \right)^{-\gamma}\exp{\left\{ -\left(\frac{r}{r_{\rm c}}\right)^{2-\gamma} \right\}},
\label{eq:sigmag}
\end{eqnarray}
where
\begin{eqnarray}
\Sigma_{0} = (2 - \gamma) \frac{M_d}{2 \pi r_c^2} \exp\left(\frac{r_\text{in}}{r_c}\right)^{2 - \gamma},
\end{eqnarray}
with $M_{\rm d}=0.2M_{\odot}$ \citep{BI20}, $r_{\rm c}=80$ au, $r_{\rm in}=8.8$ au, and $\gamma=0.2$ from fitting of dust continuum emission \citep{Kwon+15}.

The maximum dust size (multiplied by $f_{\rm fill}$) increases toward the inner regions and remains below 1 mm in the outer disk ($r \gtrsim 60$ au).  
This is because the outer disk's spectral index is $\sim3.7$ at $\lambda >$ a few mm (Figure \ref{fig:HLTau}), which suggests $a_{\rm max}f_{\rm fill}\lesssim1$ mm (Figure \ref{fig:opac}).
Figure \ref{fig:full} also compares the predicted maximum dust size with the Stokes number given by
\begin{eqnarray}
{\rm St}= \frac{\pi}{2}\frac{\rho_{\rm m}a_{\rm max}f_{\rm fill}}{\Sigma_{\rm g}},
\label{eq:St}
\end{eqnarray}
where $\rho_{\rm m}=1.675~{\rm g~cm^{-3}}$ is the dust material density.  
Beyond 60 au, ${\rm St} \lesssim 10^{-2}$. 
Toward the inner disk, the spectral index approaches 2 even at longer wavelengths.
Small grains can produce $\alpha \sim 2$ only when the optical depth is greater than unity,  
whereas large grains always result in $\alpha \sim 2$ regardless of optical depth, favoring the presence of large grains.  
In the optically thick regime, scattering can reduce intensity and alter the spectral index from $\alpha = 2$ \citep[e.g.,][]{Ueda+20}.
However, between 0.45 and 1.3 mm, the spectral index shows no significant deviation from $\alpha = 2$ (see Figure \ref{fig:alpha} in Appendix \ref{sec:spectralindex}), except in gap regions, supporting either $a_{\rm max} f_{\rm fill} \ll \lambda / 2\pi$ or $a_{\rm max} f_{\rm fill} \gg \lambda / 2\pi$.

\begin{figure}[tbh]
\begin{center}
\includegraphics[width=\columnwidth]{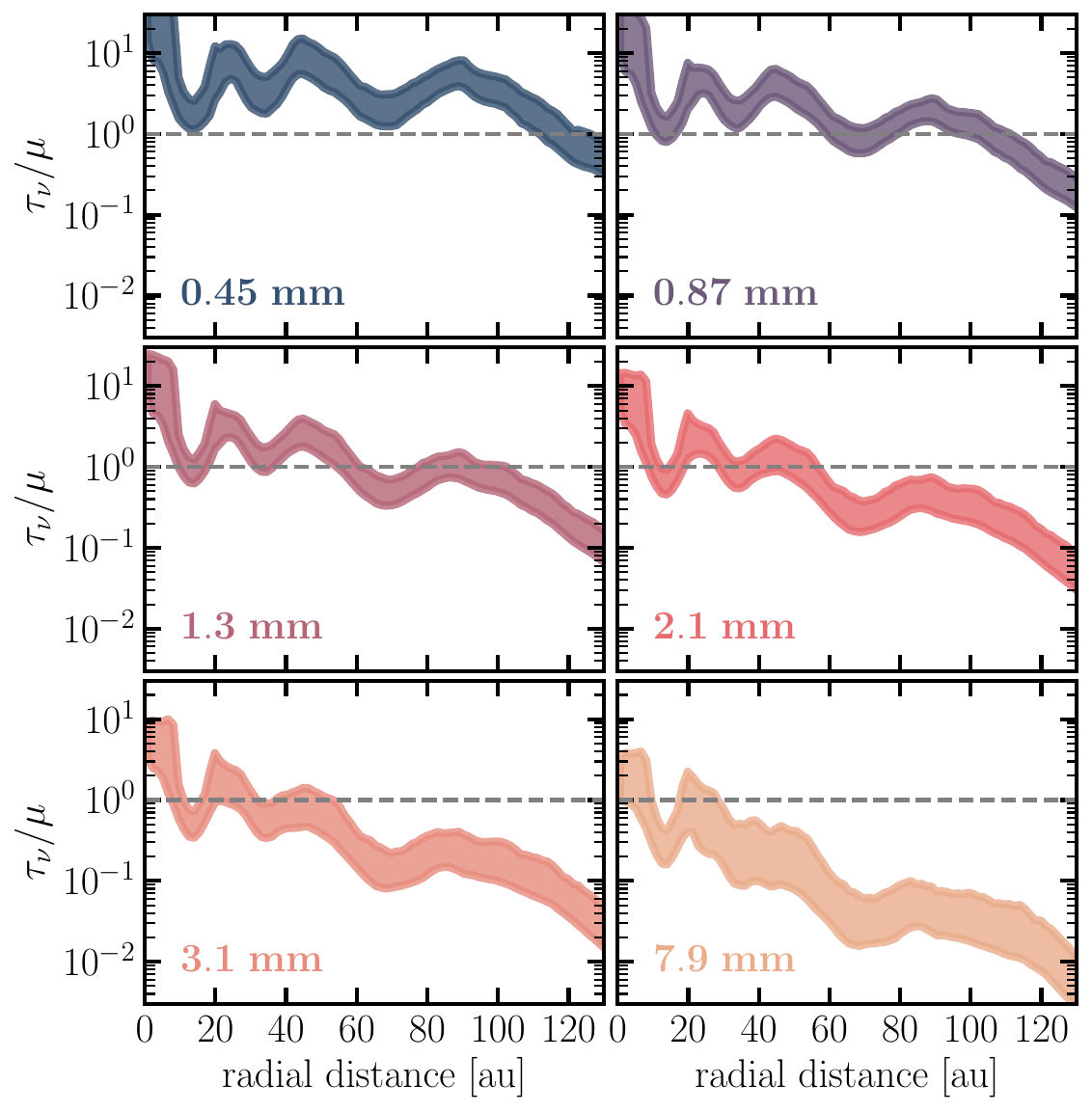}
\caption{
Line-of-sight optical depth at each observing wavelength, calculated using the 68\% confidence interval of the posterior distributions.
The horizontal dashed line marks an optical depth of unity.
}
\label{fig:tau}
\end{center}
\end{figure}

The amorphous carbon fraction also shows an increasing trend toward the inner region. 
At $r \gtrsim 60$ au, the amorphous carbon fraction is preferred to be $\lesssim0.4$ (corresponding to $40\%\times0.4=16$\% in the total dust mass), which more closely resembles the DSHARP dust model than the RICCI model.
This is because amorphous-carbon-rich dust exhibits an opacity index smaller than 1.7 at $a_{\rm max}f_{\rm fill}\lesssim100~{\rm \mu m}$, allowing it to reproduce the observed spectral index of 3.7 only when the dust is compact and $a_{\rm max}f_{\rm fill}\sim\lambda/2\pi$, at which Mie interference occurs.
In contrast, amorphous-carbon-rich dust is preferred within $\sim40$ au.
As discussed later (see Section \ref{sec:fix}), this is likely because the disk is optically thick at most observing wavelengths, and hence higher absorption opacity can explain the observed behavior across a wider range of parameters.
In this case, the preference for amorphous-carbon-rich dust arises from prior selection rather than intrinsic composition.

The preferred filling factor and power-law index of dust-size distribution have large uncertainties and are not well constrained.
The filling factor is preferred to be $\lesssim0.3$ at $>40$ au, but more compact dust appears to be still acceptable.
We examine the details in the next section.

Figure \ref{fig:tau} shows the optical depth at each wavelength, calculated using the 68\% confidence interval of the posterior distributions.
Clear drops in optical depth at $\sim13$, 30, and 70 au correspond to deep gaps in the millimeter emission.
The disk is (at least marginally) optically thick inside $\sim$60 au at $\lambda \lesssim 3.1$ mm, while the outer region is likely optically thin at $\lambda \gtrsim 2.1$ mm.

\subsection{Fixing the dust composition and porosity} \label{sec:fix}

In this section, we perform MCMC fitting with fixed dust composition and filling factor to assess their impact.  
We set $f_{\rm AC} = 0.3$ or $0.8$ to test the robustness of its radial variation seen in the full fitting.  
Similarly, $f_{\rm fill}$ is fixed at 0.1 or 0.8 to verify the preference for $f_{\rm fill} \lesssim 0.3$.  

\begin{figure*}[tbh]
\begin{center}
\includegraphics[width=2\columnwidth]{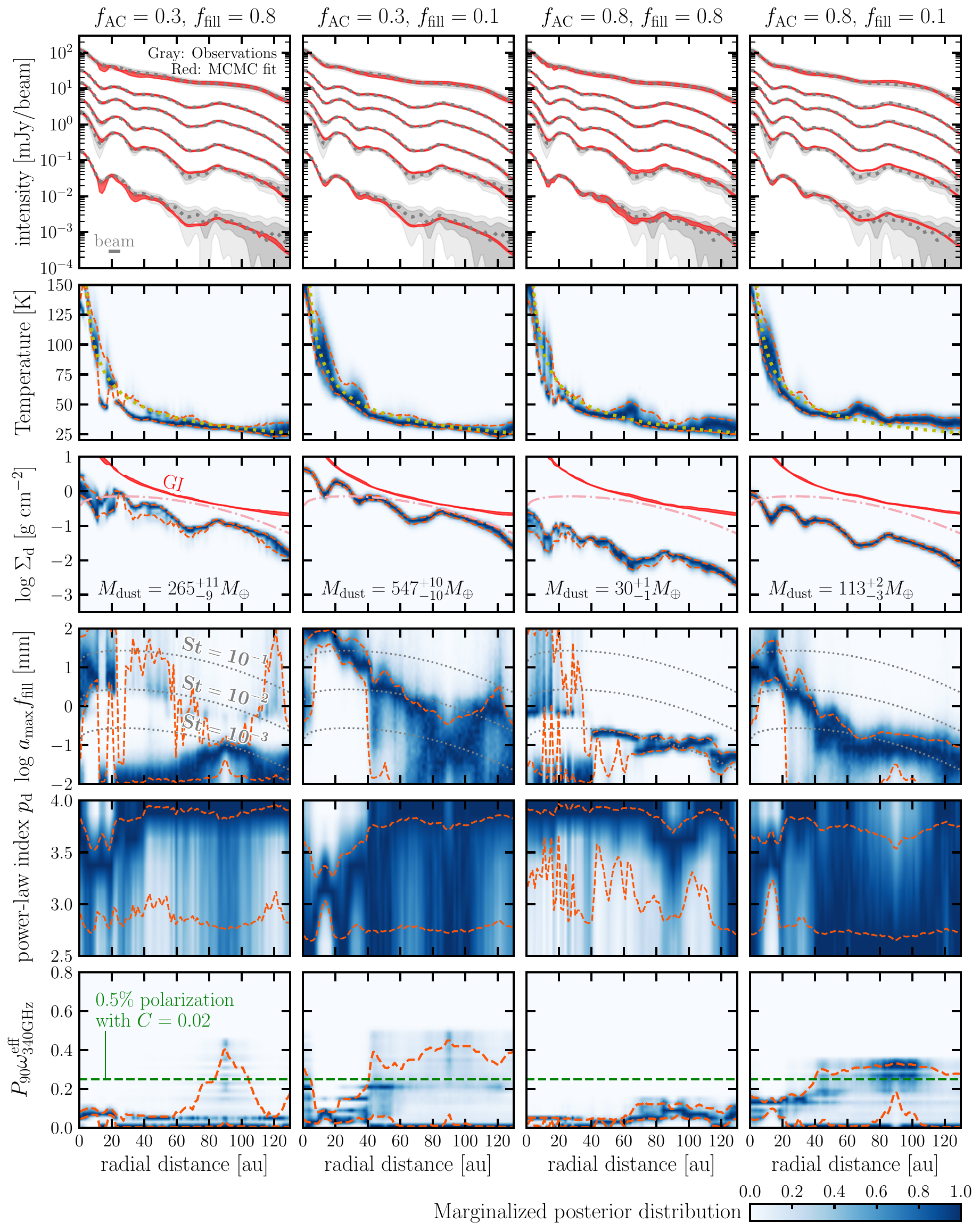}
\caption{
MCMC fitting results with  the dust composition and filling factor fixed to 
$f_{\rm AC}=0.3$ and $f_{\rm fill}=0.8$ (left), 
$f_{\rm AC}=0.3$ and $f_{\rm fill}=0.1$ (second left), 
$f_{\rm AC}=0.8$ and $f_{\rm fill}=0.8$ (second right), and 
$f_{\rm AC}=0.8$ and $f_{\rm fill}=0.1$ (right).
Top: comparison of observed intensities with MCMC-derived models, computed using the 68\% confidence interval of the posterior distributions.
Bottom five rows: 
marginalized posterior distributions of four fitting parameters, as well as the polarization efficiency at $\lambda=0.87$ mm.  
The green dashed line indicates polarization efficiency of 0.25, corresponding to 0.5\% polarization with $C=0.02$, while the other lines are defined in the same way as in Figure \ref{fig:full}.
}
\label{fig:MCMC_comparison}
\end{center}
\end{figure*}

Figure~\ref{fig:MCMC_comparison} presents the posterior distributions of four parameters from MCMC fitting with fixed $f_{\rm fill}$ and $f_{\rm AC}$, along with intensity profiles within the 68\% confidence interval.  
All models reasonably reproduce the observed profiles, though full fitting favors $f_{\rm AC} \gtrsim 0.6$ for $r \lesssim 40$ au and $f_{\rm AC} \lesssim 0.4$ for $r \gtrsim 60$ au (Figure~\ref{fig:full}).  
Similarly, while full fitting suggests $f_{\rm fill} \lesssim 0.3$, compact dust ($f_{\rm fill} = 0.8$) also fits well.
These results highlight the influence of prior distributions (usually a uniform distribution is assumed), necessitating careful interpretation.

The dust temperature profile remains similar regardless of composition and filling factor, although models with amorphous carbon tend to be slightly warmer.
The preferred dust surface density is highly sensitive to composition.  
For $f_{\rm AC} = 0.3$, it approaches that of a gravitationally unstable disk.
We note that if $f_{\rm AC}<0.3$, the dust surface density is likely larger than that of a gravitationally unstable disk, indicating $f_{\rm d2g}>0.01$.
In contrast, for $f_{\rm AC} = 0.8$, the dust surface density is an order of magnitude lower than that for $f_{\rm AC} = 0.3$.
With fixed composition, the dust surface density declines with distance, whereas full fitting yields a flatter profile.
The total dust mass within 140 au is $265_{-9}^{+11}$, $547_{-10}^{+10}$, $30_{-1}^{+1}$ and $113_{-3}^{+2}M_{\oplus}$ for the models with $f_{\rm AC} = 0.3$ and $f_{\rm fill} = 0.8$, $f_{\rm AC} = 0.3$ and $f_{\rm fill} = 0.1$, $f_{\rm AC} = 0.8$ and $f_{\rm fill} = 0.8$, and $f_{\rm AC} = 0.8$ and $f_{\rm fill} = 0.1$, respectively.

The predicted maximum dust size exhibits more complex behavior than temperature and surface density.
If the dust is organics-rich ($f_{\rm AC}=0.3$) and $f_{\rm fill}=0.8$, the MCMC fitting prefers dust with $a_{\rm max}f_{\rm fill}\lesssim100~{\rm \mu m}$ in most regions except $r\lesssim20$ au, but $a_{\rm max}f_{\rm fill}\gtrsim1~{\rm mm}$ also appears to be acceptable.
This is because the opacity index $\beta$ of compact dust exhibit a peak at $a_{\rm max}f_{\rm fill}\sim\lambda/2\pi$, resulting in two distinct solutions to explain $\alpha\sim3.7$; one where $a_{\rm max}f_{\rm fill}\ll \lambda/2\pi$ and the other where $a_{\rm max}f_{\rm fill}$ is slightly larger than $\lambda/2\pi$.
Between these two sizes, the enhancement of $\beta$ due to Mie interference makes it unlikely to explain the observed spectral index of $\lesssim3.7$.
In contrast, the organics-rich dust with $f_{\rm fill}=0.1$ can range in size between 0.01 and 1 mm beyond 40 au.
The corresponding Stokes number is as large as ${\rm St}\sim10^{-2}$.
This is because the porous dust exhibits $\alpha=3.7$ over a broader size range due to the suppression of Mie interference.

The amorphous-carbon-rich dust ($f_{\rm AC}=0.8$) shows an almost flat radial dust size profile with $a_{\rm max}f_{\rm fill}$ on the order of $100~{\rm \mu m}$ beyond 40 au.
This is because $\alpha\sim3.7$ can be reproduced by amorphous-carbon-rich dust only when $a_{\rm max}f_{\rm fill}\sim\lambda/2\pi$.
The smaller dust size, compared to moderately porous organics-rich dust, results in a smaller Stokes number of $\sim10^{-3}$.

\subsection{Scattering Polarization} \label{sec:pol}

In Sections \ref{sec:full} and \ref{sec:fix}, we focus on the Stokes I emission of the HL Tau disk.
However, the HL Tau disk is also rich in polarization data, which further constrains the dust properties \citep{Kataoka+17,Stephens+17,Stephens+23,Lin+24}.
Recent high-resolution ALMA Band 7 polarimetric observations reveal that scattering-induced polarization dominates in rings, while alignment-induced polarization prevails in gaps \citep{Stephens+23}.  
The exact contribution of each mechanism remains uncertain, but scattering polarization is likely $\gtrsim 0.5$\% at least in the rings.

Figure~\ref{fig:MCMC_comparison} also presents the posterior distribution of polarization efficiency at 0.87 mm, $P_{90} \omega^{\rm eff}_{\rm 345GHz}$.
Estimating the actual polarization degree requires numerical simulations accounting for multiple scattering and disk inclination, 
but it can be roughly approximated as $C P_{90} \omega^{\rm eff}_{\nu}$ with $C \sim 0.02$ \citep{Kataoka+16}.
The model with $f_{\rm fill} = 0.8$ and $f_{\rm AC} = 0.8$ fails to produce sufficient scattering polarization,  
suggesting that compact, amorphous-rich dust is not responsible for polarized emission in the HL Tau disk.

Based on this, we perform the MCMC analysis with an additional prior based on the polarization efficiency at 0.87 mm:
\begin{align}
& p_{2}(r,a_{\rm max},f_{\rm fill},f_{\rm AC},p_{\rm d}) \nonumber \\ 
&= \frac{1}{2} \left\{ 1+\tanh{\left(\frac{P_{\rm 90}\omega^{\rm eff}_{\rm 345GHz}-0.15}{0.1}\right)} \right\}.
\label{eq:prior-pomg}
\end{align}
This formulation yields $p_{2}\sim1$ for $P_{90}\omega^{\rm eff}_{\rm 345GHz}\gtrsim0.25$, while $p_{2}\sim0$ for $P_{90}\omega^{\rm eff}_{\rm 345GHz}\lesssim0.05$, with a transition from 0 to 1 within $0.05\lesssim P_{90}\omega^{\rm eff}_{\rm 345GHz} \lesssim 0.25$.
We note that the actual value of $P_{90}\omega^{\rm eff}_{\rm 345GHz}$ in the disk is expected to vary with radius. 
However, we adopt Equation \eqref{eq:prior-pomg} for the entire region of the disk, as a definitive profile of the scattering polarization degree has not yet been obtained.
With Equations \eqref{eq:t_prior} and \eqref{eq:prior-pomg}, the modified prior is given by $p_{1}p_{2}$.

\begin{figure*}[tbh]
\begin{center}
\includegraphics[width=2\columnwidth]{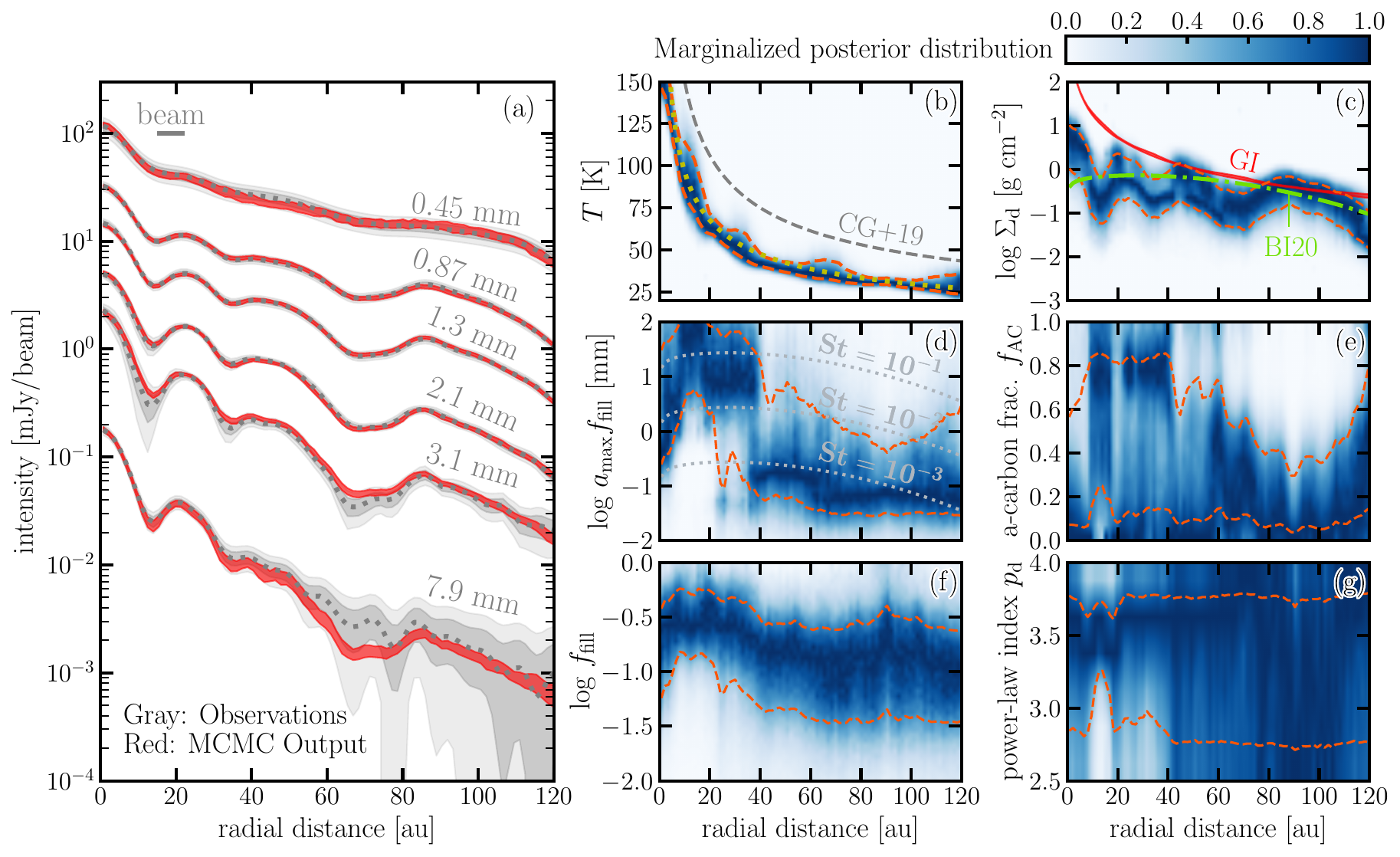}
\caption{
Same as Figure \ref{fig:full} but with adopting the prior based on the polarization efficiency at 0.87 mm (Equation \ref{eq:prior-pomg}).
}
\label{fig:full_pol}
\end{center}
\end{figure*}

Figure~\ref{fig:full_pol} presents the marginalized posterior distribution from full MCMC fitting with the modified prior based on the polarization efficiency at 0.87 mm.
Overall behavior is similar to that obtained without the prior given by Equation \eqref{eq:prior-pomg} (Figure \ref{fig:full}).
At $r \gtrsim 40$ au, the maximum dust size is likely $\sim100$ ${\rm \mu m}$, whereas the fitting without the polarization-based prior allows smaller grains.
The filling factor is preferred to be in the range $0.03 \lesssim f_{\rm fill} \lesssim 0.3$, while full fitting without the polarization-based prior allows for more porous dust.
The posterior distribution of the amorphous-carbon fraction $f_{\rm AC}$ is more uniform at $r\lesssim40$ au compared to that without the polarization-based prior, making organics-rich dust also a plausible solution due to its higher polarization efficiency.

\section{Discussion}\label{sec:discussion}

\subsection{Dust size and mass in the HL Tau disk}
The dust properties of the HL Tau disk have been studied with multi-wavelength analysis.
\citet{Carrasco-Gonzalez+19} inferred mm-sized dust throughout the disk from 0.9–8 mm data, while \citet{Guerra-Alvarado+24}, using 0.45 mm data, found smaller grains ($\sim0.3$-1 mm) beyond 60 au. 
Our MCMC analysis suggests that $a_{\rm max}f_{\rm fill}\sim 100~{\rm \mu m}$ at $\gtrsim40$ au, consistent with the porous dust model in \citet{Guerra-Alvarado+24}.
For compact dust, we prefer even smaller grains beyond 40 au.
This discrepancy is mainly due to filtering effects in the ALMA Band 4 data, where fainter emission steepens the spectral index between 1.3 and 2.1 mm, favoring larger dust. 
Furthermore, our new 3.1 mm data gives $\alpha\approx3.7$ between 2.1 and 3.1 mm, supporting the small-grain solution. 
Dust mass estimates also vary: \citet{Carrasco-Gonzalez+19} derived $300M_{\oplus}$ for compact dust (similar to our estimate with $f_{\rm AC}=0.3$ and $f_{\rm fill}=0.8$), whereas \citet{Guerra-Alvarado+24} reported 600 and $2000M_{\oplus}$ for compact and porous dust with the DSHARP composition, respectively. 
With $f_{\rm fill}=0.1$, which is favored by polarization, $f_{\rm AC}\lesssim0.3$ yields masses exceeding that of a gravitational unstable disk, and hence $f_{\rm AC} \gtrsim0.3$ is preferable.
The total mass is estimated at $547M_{\oplus}$ for $f_{\rm AC}=0.3$ and $113M_{\oplus}$ for $f_{\rm AC}=0.8$ with $f_{\rm fill}=0.1$.

The presence of $100~{\rm \mu m}$-sized dust is consistent with the prediction by \cite{OT19}, who argue that the dust in the HL Tau disk is fragile, with a critical fragmentation velocity --- above which dust collisions result in fragmentation --- of $0.5~{\rm m~s^{-1}}$ (see also \citealt{Jiang+24,Ueda+24} for other arguments in support of the fragile dust scenario).  
\cite{OT19} suggested that the fragility could be attributed to non-sticky ${\rm CO_{2}}$ ice mantles \citep{Musiolik+16}, such that dust grows up to $\sim1$ cm within the ${\rm CO_{2}}$ snowline at $T\sim60~{\rm K}$ ($r\sim25$ au in our model), leading to a significant drop in dust surface density within the ${\rm CO_{2}}$ snowline due to radial drift.
Our results also indicate the presence of cm-sized dust in similar inner regions, but it is unclear if the size increase is associated with the ${\rm CO_{2}}$ snowline because of large uncertainties. 
The predicted dust surface density shows a drop within 20 au, which could be related to a change in drift velocity caused by ${\rm CO_{2}}$ sublimation.

\subsection{Dust porosity in the HL Tau disk}
The preference for the moderately porous dust ($0.03\lesssim f_{\rm fill} \lesssim 0.3$) is consistent with the modeling by \cite{Zhang+23}, in which the Stokes $I$ and polarized emission are interpreted with the DSHARP dust model.  
More recent modeling on multi-wavelength polarimetric observations of the HL Tau disk showed a slow decrease of the scattering component with increasing wavelength, again supporting the porous dust solution \citep{Lin+24}.
Similar conclusions have been provided for the IM Lup disk \citep{Ueda+24}.
The dust filling factor achieved by simple pair-wise collisional growth is expected to be significantly smaller than 0.01 (e.g., \citealt{Suyama+08}), which is not consistent with the results of this study. 
In contrast, models in which voids are filled by small fragments produced via collisional fragmentation yield filling factor around 0.1, which is broadly consistent with our findings \citep{Dominik+16,Tanaka+23}. 
Although this study assumes that the filling factor is constant across the dust size distribution, filling factor may, in reality, depend on dust size. 
Recent near-infrared observations suggest that small grains in disk surface layer may be highly porous \citep{Ginski+23,Tazaki+23}. 
Our results are more sensitive to the filling factor of the upper end of the size distribution that contributes to the millimeter emission.

\subsection{Dust composition in the HL Tau disk}
The full fitting with variable dust composition shows a preference for amorphous carbon-rich dust within 40 au. Refractory organics are thermally decomposed at temperatures of $\sim$300--500K \citep{Nakano+03}, whereas amorphous carbon can survive at much higher temperatures ($>1000$K; \citealt{Gail01}). 
This suggests that dust within 40 au may have been heated to temperatures exceeding 300--500K, sufficient to decompose refractory organics while leaving amorphous carbon intact. 
The presence of thermally processed dust at large radii could be attributed to an accretion outburst, which causes a transient increase in the disk temperature \citep{van'tHoff+18,Tobin+23,HK23,Calahan+24,Colmenares+24}, or to the outward diffusion of dust from the warmer inner regions \citep{Gail01,KG04,Ciesla07,Ciesla09,Zhou+22}. 
However, the apparent preference for amorphous carbon-rich dust within 40 au of the disk is not conclusive in this study, and further investigation is needed.

If the dust composition is uniform across the disk, given that the gas disk is near gravitational instability, dust with $f_{\rm AC} \sim 0.3$ is preferred under a typical dust-to-gas mass ratio of 0.01.
However, a hypothesis that the HL Tau disk is already depleted in dust due to radial drift,  consistent with the low dust surface density in the model with amorphous-carbon-rich dust, cannot be ruled out.

ALMA Band 7 polarization data suggest the dust is unlikely to be both compact and amorphous-carbon-rich. 
In contrast, disk population synthesis models favor such dust to explain the low spectral index and large flux of Class II disks \citep{Delussu+24}.
The tension between these findings suggests that dust composition varies across different disks and/or that additional mechanisms sustain the dust mass to maintain a high optical depth required to explain the observations.

\subsection{Planet formation in the HL Tau disk}

\begin{figure*}[tbh]
\begin{center}
\includegraphics[width=2\columnwidth]{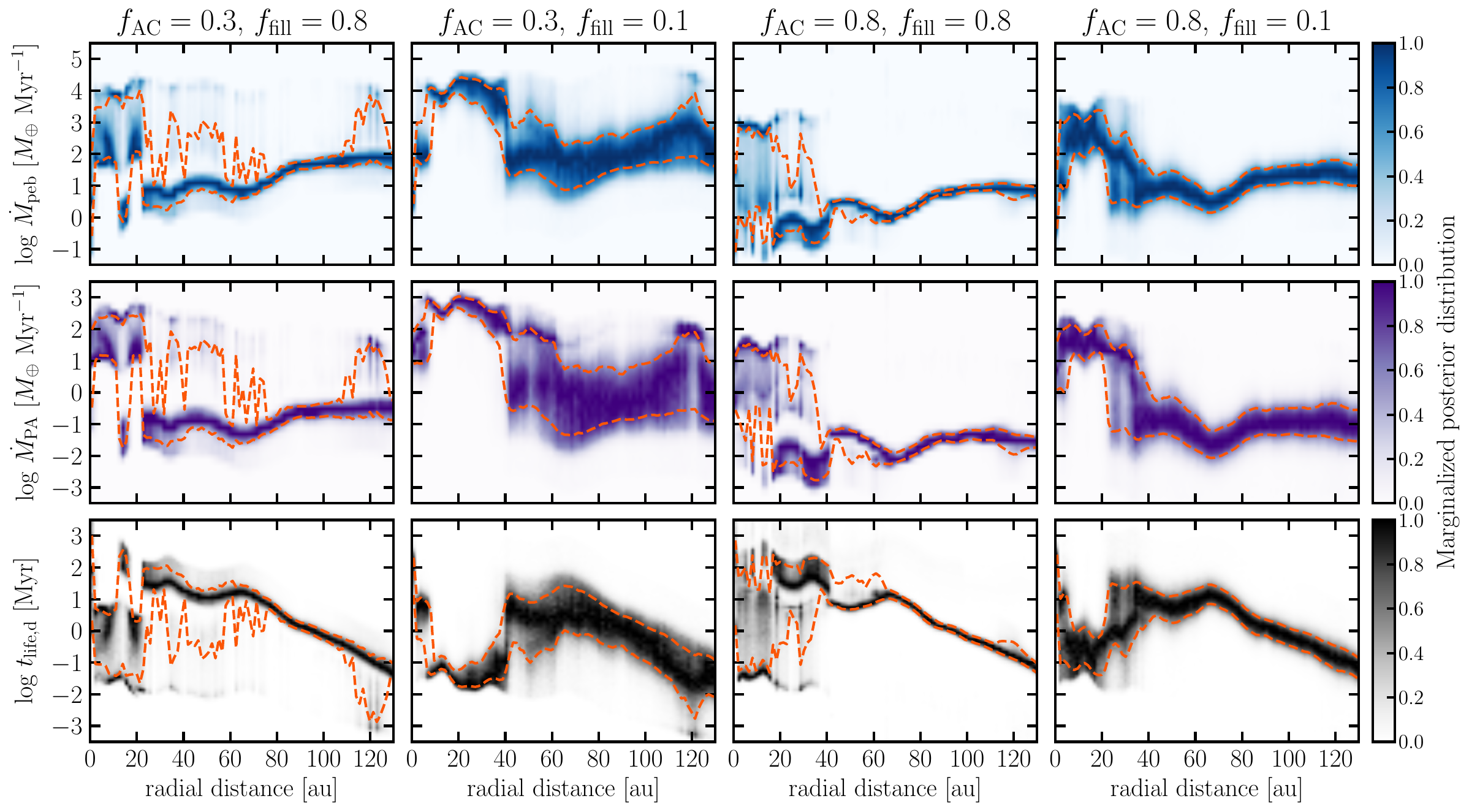}
\caption{
Marginalized posterior distributions of the radial pebble flux (top), pebble accretion rate onto a $3M_{\oplus}$ planet (middle), pebble lifetime (bottom) derived from the posterior distributions based on the polarization-based prior. 
}
\label{fig:pebble}
\end{center}
\end{figure*}

In this section, we discuss the possibility of planet formation via pebble accretion (e.g., \citealt{LJ14}) in the HL Tau disk based on our dust properties estimation.

The radial pebble flux is calculated as 
\begin{eqnarray}
\dot{M}_{\rm peb}= 2\pi r v_{\rm r} \Sigma_{\rm d},
\label{eq:Mdot_pebble}
\end{eqnarray}
with $v_{\rm r}$ being the radial drift velocity of dust defined as \citep{Adachi+76}
\begin{eqnarray}
v_{\rm r}= -2\frac{\rm St}{1+{\rm St}^{2}}\eta v_{\rm K},
\label{eq:vr}
\end{eqnarray}
where $v_{\rm K}=r\Omega_{\rm K}$ is the Keplerian velocity and $\eta$ is the deviation of the rotational velocity from the Keplerian velocity defined as 
\begin{eqnarray}
\eta = -\frac{1}{2} \left( \frac{c_{\rm s}}{v_{\rm K}} \right)^{2} \frac{\partial \ln \rho_{\rm g}c_{\rm s}^{2}}{\partial \ln r},
\label{eq:eta}
\end{eqnarray}
where $\rho_{\rm g}=\Sigma_{\rm g}/\sqrt{2\pi}h_{\rm g}$ with $h_{\rm g}=c_{\rm s}/\Omega_{\rm K}$ being the gas scale height.
We adopt Equation \eqref{eq:temp_simple} to calculate $c_{\rm s}$.
We note that we assume a smooth profile for the gas disk structure, whereas $\eta$ depends on the radial gradients of gas disk properties, which can be sensitive to the presence of substructures.

Given the radial pebble flux defined by Equation \eqref{eq:Mdot_pebble}, the pebble accretion rate onto a seed planet is given by
\begin{eqnarray}
\dot{M}_{\rm PA} = P_{\rm eff}\dot{M}_{\rm peb},
\label{eq:PA}
\end{eqnarray}
where $P_{\rm eff}$ is the pebble accretion efficiency, representing the fraction of drifting pebbles that are accreted by the planet.
The pebble accretion efficiency is defined as 
\begin{eqnarray}
P_{\rm eff}= \min(P_{\rm eff,2D},P_{\rm eff,3D}),
\label{eq:Peff}
\end{eqnarray}
where $P_{\rm eff,2D}$ and $P_{\rm eff,3D}$ represent the pebble accretion efficiency in the 2D and 3D regimes, respectively, corresponding to cases where the pebble accretion radius is larger than or smaller than the pebble scale height \citep{LO18}:
\begin{eqnarray}
P_{\rm eff,2D}= 0.32 \sqrt{\frac{q_{\rm p}}{{\rm St}\eta^{2}} \frac{\Delta v}{v_{\rm K}}},
\label{eq:Peff2D}
\end{eqnarray}
and 
\begin{eqnarray}
P_{\rm eff,3D}= 0.39 \frac{q_{\rm p}r}{\eta h_{\rm p}},
\label{eq:Peff3D}
\end{eqnarray}
with $q_{\rm p}$, $\Delta v$ and $h_{\rm p}$ being the planet-to-star mass ratio, relative velocity between planet and dust and pebble scale height, respectively.
The pebble scale height is defined as \citep{YL07}
\begin{eqnarray}
h_{\rm p} = h_{\rm g} \left(1+ \frac{\rm St}{\alpha_{\rm vert}} \frac{1+2{\rm St}}{1+{\rm St}} \right)^{-1/2},
\end{eqnarray}
with $\alpha_{\rm vert}$ being the turbulence parameter in the vertical direction.
We assume $\alpha_{\rm vert}=10^{-4}$.
We also calculate the pebble accretion rate for $\alpha_{\rm vert} = 10^{-3}$ and $10^{-5}$, and find that the accretion rate is $\sim3$ times lower for $\alpha_{\rm vert} = 10^{-3}$ and $\sim3$ times higher for $\alpha_{\rm vert} = 10^{-5}$, compared to the case with $\alpha_{\rm vert} = 10^{-4}$.
The relative velocity between planet and dust consists of contributions from the headwind and the Keplerian shear, and is given by \citep{LO18}
\begin{eqnarray}
\Delta v/v_{\rm K} = \left( 1+ \frac{5.7q_{\rm p}{\rm St}}{\eta^{3}}\right)^{-1} \eta + 0.52 (q_{\rm p} {\rm St})^{1/3}.
\end{eqnarray}

Figure \ref{fig:pebble} shows the posterior distributions of the radial pebble flux, pebble accretion rate onto a planet with mass of $3M_{\oplus}$, and the pebble lifetime $t_{\rm life,d}$ obtained from the MCMC analysis with the polarization-based priors.
The pebble lifetime at $r$ is defined as the total amount of dust residing at $>r$, $M_{\rm dust}(>r)$, divided by the pebble flux at $r$.
One important finding is that no significant decrease in dust flux is observed inside the gaps. 
This is because both the dust surface density and grain size do not vary significantly between the inner and outer regions of the gap.
This suggests that the gaps are leaky, and that the gap outer edge does not trap pebbles efficiently enough to cause a substantial depletion of dust in the inner regions (e.g., \citealt{Pinilla+21,Stammler+23,VanClepper+25}).
\citet{Eriksson+20} argued that, if the dust size is around 100 ${\rm \mu m}$ and the planet has not yet reached the pebble isolation mass, dust can persist in the inner region of the gap over long timescales, which is consistent with our results.

The predicted radial pebble flux is confined to a narrow parameter space in compact dust models due to the small uncertainty in the maximum dust size, whereas it exhibits greater uncertainty in porous dust models due to the larger uncertainty in the maximum dust size (see Figure \ref{fig:MCMC_comparison}).
Beyond 40 au, the radial pebble flux is predicted to be $\lesssim10^{2}M_{\oplus}~{\rm Myr^{-1}}$ for most models, except for the porous organics-rich dust model ($f_{\rm AC}=0.3$ and $f_{\rm fill}=0.1$), where it can be as high as $10^{3}M_{\oplus}~{\rm Myr^{-1}}$.
The large pebble flux in the porous organics-rich dust model is due to the combined effects of the predicted large dust size and high dust surface density.

The estimation of pebble flux in protoplanetary disks remains a non-trivial task, owing to significant uncertainties in dust size, dust mass, and gas disk profile.
\citet{Zhang+20} estimated a pebble flux of 15–60~$M_\oplus$ Myr$^{-1}$ based on the measured C/H ratio inside 70 au in the HD163296 disk.
\citet{Romero-Mirza+24}, using JWST observations of water vapor, inferred pebble fluxes of 30–370~$M_\oplus$ Myr$^{-1}$ at the water snow line in six out of seven disks.
These values are consistent with the pebble fluxes of our organics-rich dust model at $>40$ au where the derived dust properties are not significantly affected by high optical thickness.
In contrast, the compact and amorphous-carbon-rich dust model yields lower pebble flux ($\lesssim10M_{\oplus}~{\rm Myr^{-1}}$) than these estimates.
The pebble mass flux in disks has also been studied using numerical models (e.g., \citealt{Krijt+18, Kalyaan+21, Drazkowska+21}).
\citet{LJ14} constructed an analytical model that predicts a flux of 95~$M_\oplus$~Myr$^{-1}$ at 1 Myr, which is consistent with our porous organics-rich dust model.
\citet{Drazkowska+21} explored how the pebble flux depends on disk parameters, showing that smaller pebbles (e.g., those resulting from a lower fragmentation velocity) are depleted more slowly, sustaining a pebble flux over time.
The combination of small dust size and high pebble flux in our porous organics-rich dust model is consistent with such numerical models that assume low fragmentation velocity (e.g., $v_{\rm frag} \sim 1$m s$^{-1}$).
On the other hand, our compact amorphous-carbon-rich dust model, which shows substantially lower pebble flux, may be more consistent with scenarios in which the pebble flux significantly decreases early due to rapid radial drift facilitated by a high fragmentation velocity.

The estimated pebble flux in the outer disk can also be used to infer the potential for inner planet formation \citep{WK25}.
\citet{Lambrechts+19} showed that cumulative pebble masses of 110 and 190 $M_\oplus$ are required to form Earth-like and Super-Earth-like planets, respectively.
Assuming that the pebble flux at terrestrial orbits is similar to that at 40 au, these cumulative pebble masses can be quickly achieved if the dust is porous and organics-rich.
However, if the dust is compact and lacks organics, the current pebble flux would need to be sustained for $\sim10$ Myr to reach a cumulative mass on the order of 100 $M_\oplus$.

Because of the large pebble flux, the pebble accretion rate can reach $10M_{\oplus}~{\rm Myr^{-1}}$ (or even higher, depending on the radial location) in the porous organics-rich dust model, allowing a giant planet core to form within 1 Myr.
In contrast, it is unlikely to form a core of giant planet from a $3M_{\oplus}$ seed within the age of HL Tau ($\sim1$ Myr) via pebble accretion at $\gtrsim40$ au if the dust is compact and/or amorphous carbon rich.
We note that the pebble accretion rate depends on the seed mass and it needs to be $\sim1M_{\oplus}$ or larger for the pebble accretion rate to exceed $1M_{\oplus}~{\rm Myr^{-1}}$ in our model.

Inside 40 au, the radial pebble flux can reach $\sim10^{4} M_{\oplus}~{\rm Myr^{-1}}$. 
However, given this high pebble flux, the inner disk would be depleted of dust within the age of HL Tau (\(\sim 1\)~Myr) because $t_{\rm life,d}<1$ Myr. 
This indicates that the maximum dust size within 40 au is overestimated probably because of high optical depth, or that the ${\rm ^{13}C^{17}O}$-based gas surface density is underestimated, causing larger ${\rm St}$.
The pebble lifetime is also $<1$ Myr at $r\gtrsim100$ au, possibly because our fitting excludes the low signal-to-noise regions beyond 140 au.

We note that direct planet formation via gravitational instability is also a viable scenario for potential planets in the observed gaps, as the gas surface density is likely close to the gravitational instability threshold at least in $50\lesssim r \lesssim 100$ au \citep{BI20}.

\subsection{Prediction for longer wavelengths observations}
Our analysis shows that the current dataset can be explained by either amorphous-carbon-rich or organics-rich dust models, emphasizing the need for future observations particularly at longer wavelengths where the disk is optically thin.

\begin{figure}[tbh]
\begin{center}
\includegraphics[width=\columnwidth]{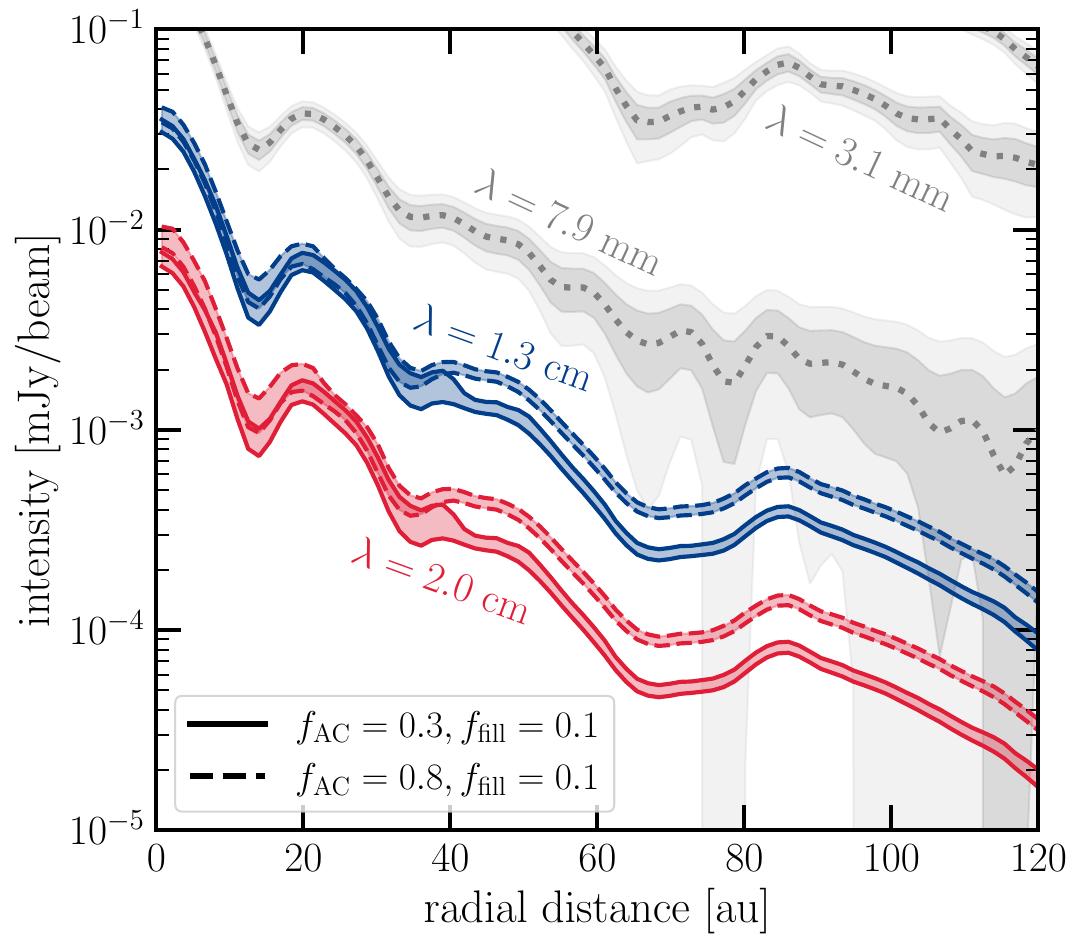}
\caption{
68\% confidence interval of the intensity at 1.3 (blue) and 2.0 cm (red) estimated from the posterior distributions obtained from the MCMC analysis.
The transparent regions with solid or dashed lines denote the models with $f_{\rm AC}=0.3$ and 0.8, respectively.
}
\label{fig:predict}
\end{center}
\end{figure}

Figure \ref{fig:predict} shows the 68\% confidence interval of the predicted intensity profiles at 1.3 and 2.0 cm, estimated from the posterior distributions sampled by the MCMC analysis.
It can be seen that the model with amorphous-carbon-rich dust ($f_{\rm AC} = 0.8$) predicts the intensity about twice that of the $f_{\rm AC} = 0.3$ model beyond 40 au.
At 1.3 cm, the difference in the two models is $\sim0.5$ and $\sim0.2~{\rm \mu Jy/beam}$ at 50 and 100 au, respectively.
Given the expected sensitivity of the Next Generation Very Large Array ($\sim0.2~{\rm \mu Jy}$ for 1 hour integration), we anticipate that the two models can be distinguished with 3$\sigma$ level by $\sim1.5$ and $\sim9$ hours of integration time. 
At 2.0 cm, longer integration times are required: around 4 hours for the emission at 50 au and 225 hours for the emission at 100 au.

\section{Summary}\label{sec:summary}
We conducted a comprehensive analysis of the HL Tau disk by modeling its radial intensity profiles across six wavelengths. Using the MCMC approach, we explored various dust properties, including temperature, surface density, maximum dust radius, composition, filling factor, and size distribution. 
Our findings are summarized as follows:

\begin{enumerate}
\item The MCMC analysis favors moderately porous, organics-rich dust ($f_{\rm AC} \lesssim 0.4$) beyond 60 au, where the spectral index reaches $\sim3.7$ (Section \ref{sec:full}), but amorphous-carbon-rich dust also reasonably reproduces the observations (Section \ref{sec:fix}).
\item The maximum dust size increases inward. Beyond 40 au, it is likely an order of $\sim100~{\rm \mu m}$ if dust is compact and/or amorphous-carbon rich, but mm-sized dust is also acceptable if the dust is moderately porous and organics-rich (Figure \ref{fig:MCMC_comparison}).
\item The inferred dust surface density is consistent with the gravitational instability threshold at $50\lesssim r \lesssim 100$ au if the dust is moderately porous and organics-rich (Figure \ref{fig:MCMC_comparison}).
The high dust surface density, combined with the potential presence of mm-sized dust, enables a seed planet to grow into a giant planet core within 1 Myr through pebble accretion (Figure \ref{fig:pebble}).
\item If the dust is amorphous-carbon-rich, the dust surface density must be lower than that estimated for organics-rich dust, suggesting that pebble accretion is unlikely to be the origin of potential planets beyond 40 au (Figure \ref{fig:pebble}).
\item Longer wavelength observations, particularly at 1.3 and 2.0 cm, could distinguish between these dust models (Figure \ref{fig:predict}).
\end{enumerate}

Our findings highlight the sensitivity of dust property estimates to the assumed dust model and underscore their significant implications for planet formation in protoplanetary disks.

\acknowledgments
This paper makes use of the following ALMA data: ADS/JAO.ALMA\#2011.0.00015.SV and
ADS/JAO.ALMA\#2019.1.00134.S.
ALMA is a partnership of ESO (representing its member states), NSF (USA) and NINS (Japan), together with NRC (Canada), MOST and ASIAA (Taiwan), and KASI (Republic of Korea), in cooperation with the Republic of Chile. 
The Joint ALMA Observatory is operated by ESO, AUI/NRAO and NAOJ. 
Numerical computations were carried out on the Smithsonian High Performance Cluster (SI/HPC), Smithsonian Institution (https://doi.org/10.25572/SIHPC) and computing facilities at the North American ALMA Science Center, the National Radio Astronomy Observatory.
The National Radio Astronomy Observatory is a facility of the U.S. National Science Foundation operated under cooperative agreement by Associated Universities, Inc.
T.U. acknowledges the support from the JSPS Overseas Research Fellowship and the Smithsonian Astrophysical Observatory.
C.C.-G. acknowledges support from UNAM DGAPA-PAPIIT grant IG101224 and from CONAHCyT Ciencia de Frontera project ID 86372.
This work was supported by JSPS KAKENHI Grant Numbers JP22K03680.

\appendix

\section{Effect of short-baseline data on the Band 4 image}

\begin{figure}[tbh]
\begin{center}
\includegraphics[width=\columnwidth]{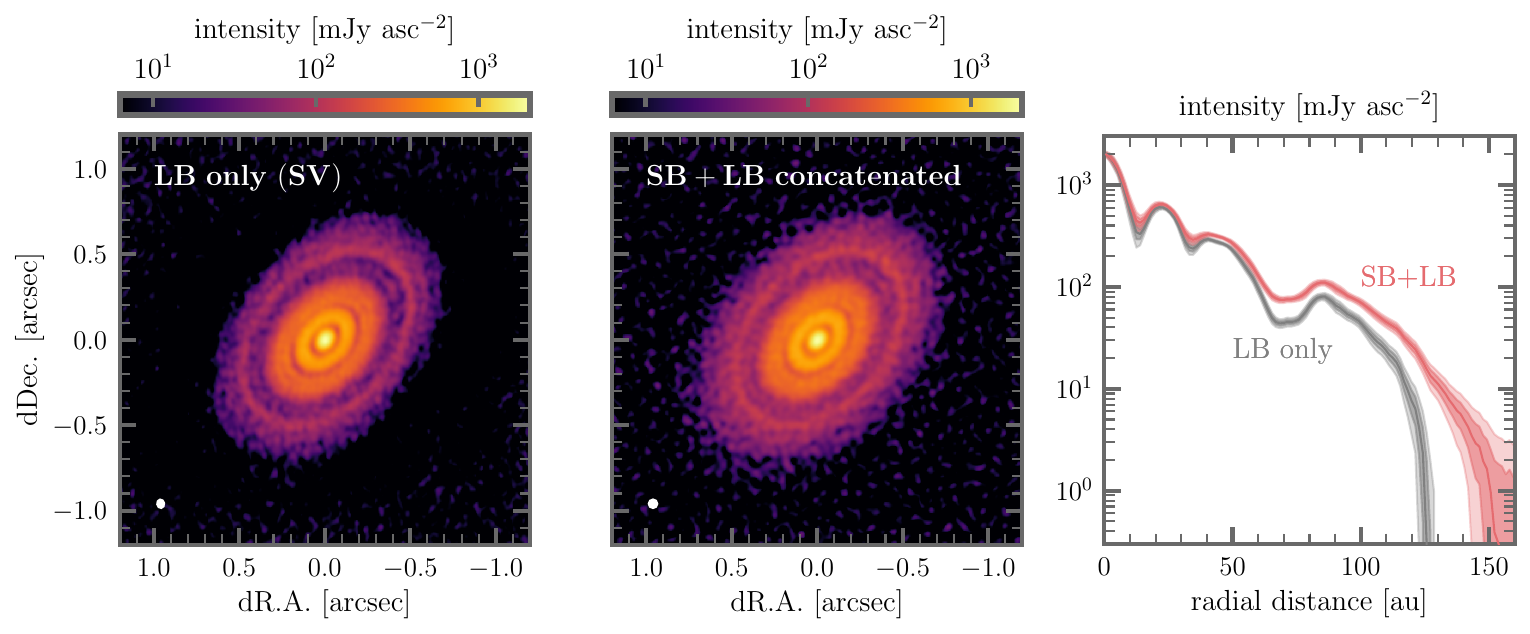}
\caption{
Comparison of the 2.1 mm intensity map from long-baseline ALMA Science Verification data (left) and our concatenated data (center).
Right: Azimuthally averaged radial intensity profiles obtained from the concatenated (red) and long-baseline data (black), respectively.
The thin and thick transparent regions indicate 1$\sigma$ and 2$\sigma$ RMS levels.
}
\label{fig:B4}
\end{center}
\end{figure}

Figure \ref{fig:B4} compares the 2.1 mm intensity of the HL Tau disk from ALMA Science Verification data and our concatenated dataset. 
The former shows a steep decline beyond 120 au, while the latter recovers the emission without significant filtering.
Beyond $\sim60$ au, the concatenated data are about twice as bright.
As shown in Figure \ref{fig:HLTau}, the spectral index at $\lambda\gtrsim2$ mm is $\sim3.7$ in the outer region. The lower intensity in the long-baseline-only data artificially increases $\alpha$ ($>3.7$), favoring compact dust with $a_{\rm max} \sim \lambda/2\pi$ due to Mie interference.
\label{sec:B4}

\section{Radial profiles of spectral index} \label{sec:spectralindex}

\begin{figure*}[tbh]
\begin{center}
\includegraphics[width=1\textwidth]{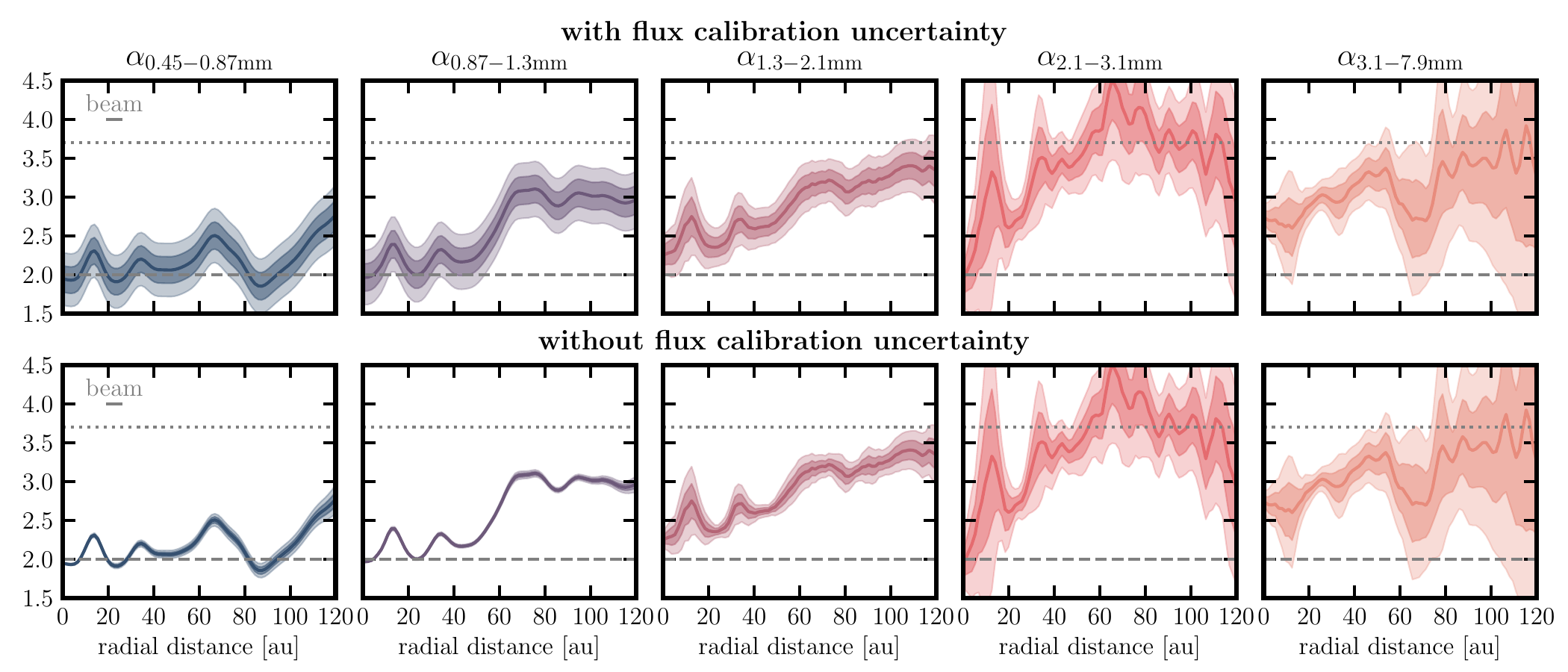}
\caption{
Radial spectral index profiles between the closest observing wavelengths, calculated with (top) and without (bottom) flux calibration uncertainty.
The thick and thin transparent regions indicate $1\sigma$ and $2\sigma$ uncertainties, respectively.
}
\label{fig:alpha}
\end{center}
\end{figure*}

Figure \ref{fig:alpha} shows the radial profiles of the spectral index between the closest observing wavelengths in our data.
The spectral index increases toward the outer region across all wavelength pairs.
At $\lambda \lesssim 2$ mm, uncertainties are dominated by flux calibration, while at $\lambda \gtrsim 2$ mm, RMS noise is the primary source.
Within 100 au, the spectral index is $\sim2$ between 0.45 and 0.87 mm, with slight increases in gap regions. 
A similar trend is seen between 0.87 and 1.3 mm, with reaching $\sim3$ beyond 60 au.
At $\lambda \gtrsim 2$ mm, the index remains $>2$ except within 20 au. 
Beyond 40 au, the spectral index is $\sim3.7$ between 2.1 and 3.1 mm, and 3.1 and 7.9 mm, though the latter has large uncertainties.

\section{Corner plot of the posterior distributions} \label{sec:corner}
\begin{figure*}[tbh]
\begin{center}
\includegraphics[width=1\textwidth]{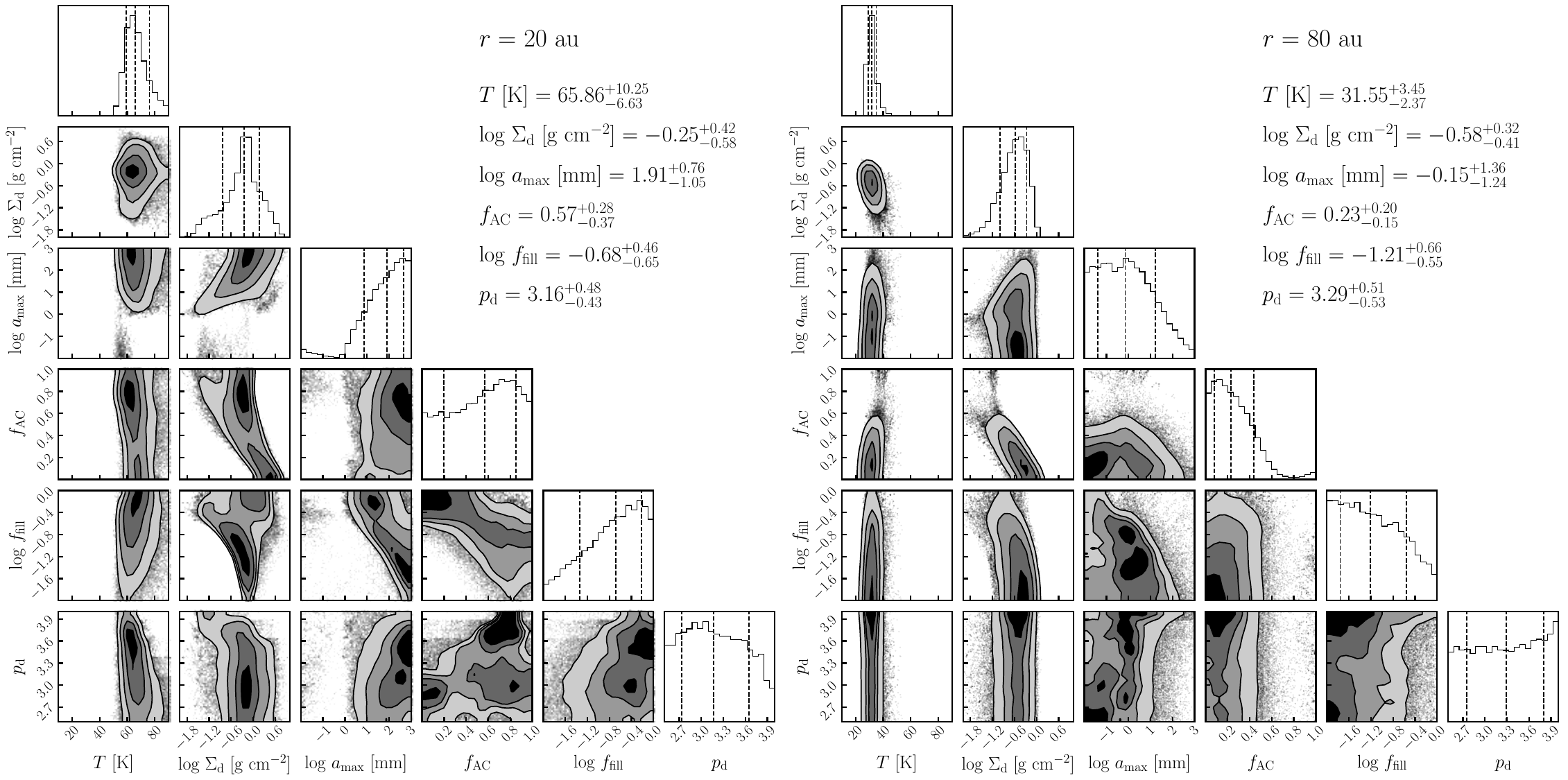}
\caption{
Corner plots showing the posterior distributions of six fitting parameters at 20 (left) and 80 au (right) obtained from the MCMC analysis with all parameters free.
The median values and 68\% confidence intervals for each parameter are indicated in the upper right of each panel.
The contours represent the 1$\sigma$, 2$\sigma$, and 3$\sigma$ confidence levels.
}
\label{fig:corner}
\end{center}
\end{figure*}

Figure \ref{fig:corner} presents corner plots of the posterior distributions for six fitting parameters at 20 and 80 au, derived from the MCMC analysis with all parameters free. At both radii, the dust temperature is nearly uncorrelated with the other parameters. The dust surface density and amorphous-carbon fraction show a diagonal trend, where a higher dust surface density corresponds to a lower amorphous-carbon fraction.
There is also a diagonal correlation between $a_{\rm max}$ and $f_{\rm fill}$, reflecting that optical properties depend primarily on the product $a_{\rm max}f_{\rm fill}$.
At 20 au, a higher filling factor is associated with a lower amorphous-carbon fraction, possibly because more compact dust exhibits higher absorption opacity due to Mie interference.
This trend does not appear at 80 au, where $a_{\rm max}$ is sufficiently smaller than the Mie regime.
At 80 au, the power-law index of the dust-size distribution $p_{\rm d}$ remains broadly distributed. 
In this region, $a_{\rm max}f_{\rm fill}\ll \lambda/2\pi$, making the opacity insensitive to $p_{\rm d}$.
In contrast, at 20 au, a larger $p_{\rm d}$ appears to correlate with larger values of $f_{\rm fill}$ and $a_{\rm max}$.
This indicate that a larger $p_{\rm d}$ --- a steeper size distribution with more small grains --- makes $a_{\rm max}f_{\rm fill}$ effectively smaller.

\bibliographystyle{aasjournal}
\bibliography{reference}

\end{document}